\pdfoutput=1
\documentclass[a4paper,12pt]{article}
\usepackage[margin=3cm,footskip=1cm]{geometry}

\usepackage[T1]{fontenc}
\usepackage[utf8]{inputenc}

\usepackage{amsmath,amssymb,amsthm,bm,mathtools,xspace,booktabs,url}
\usepackage{graphicx,etoolbox}

\usepackage[shortlabels]{enumitem}
\setlist{
  listparindent=\parindent,
  parsep=0pt,
}

\SetEnumitemKey{:alph}{
  label=\textup{(\alph*)}
}
\SetEnumitemKey{:roman}{
  label=\textup{(\roman*)}
}

\setlist[enumerate]{
  :alph
}

\AtBeginEnvironment{thebibliography}{%
  \small
  \setlength\itemsep{0.75em plus 0.5em minus 0.75em}%
}

\usepackage{caption}
\usepackage[raggedright]{titlesec}

\numberwithin{equation}{section} 

\theoremstyle{plain} 

\theoremstyle{definition} 

\usepackage{authblk}

\makeatletter
\newcommand\CorrespondingAuthor[1]{%
  \begingroup%
  \def\@makefnmark{}%
  \footnotetext{Corresponding author: #1}%
  \endgroup%
}
\makeatother

\makeatletter
\renewenvironment{abstract}{%
  \small%
  \providecommand\keywords{%
    \par\medskip\noindent\textit{Keywords:}\xspace}%
  \begin{center}%
    \bfseries \abstractname\vspace{-.5em}\vspace{\z@}%
  \end{center}%
  \quote%
}{\endquote}
\makeatother

\usepackage[above,below,section]{placeins}

\usepackage[load-configurations=abbreviations]{siunitx}

\usepackage[all]{onlyamsmath}

\usepackage{lipsum}

\DeclarePairedDelimiter\norm\lVert\rVert
\usepackage{xcolor}
\usepackage[draft]{fixme}
\fxsetup{ layout =marginnote, marginface=\scriptsize }

\newcommand\be{\mathbf{e}}
\newcommand\bu{\bm{u}}
\newcommand\bv{\bm{v}}
\newcommand\bx{\bm{x}}
\newcommand\by{\bm{y}}

\newcommand\bX{\bm{X}}
\newcommand\bY{\bm{Y}}
\newcommand\bZ{\bm{Z}}

\newcommand\R{\mathbb{R}}
\newcommand\mean{\mathrm E}
\newcommand{\var}{\mathrm{Var}}

\usepackage{fixltx2e}

\graphicspath{{.}{./code/pix-auto/}}

\usepackage{xparse,amsmath}

\ExplSyntaxOn
\clist_new:N \l_dlf_clist
\NewDocumentCommand\tsub{ m }{
  \sb{
    \clist_clear:N \l_dlf_clist
    \clist_set:Nn \l_dlf_clist {#1}
    \clist_pop:NN \l_dlf_clist \l_tmp
    \textup{\l_tmp}
    \clist_if_empty:NF \l_dlf_clist{
      ,
      \clist_use:Nn \l_dlf_clist{,}
    }
  }
}
\ExplSyntaxOff

\NewDocumentCommand\For{s m}{
  \IfBooleanTF{#1}{\quad}{\qquad}
  \text{#2}
}

\begin{document}
\title{Functional summary statistics
  for point processes on the sphere with an application to
  determinantal point processes}

\author{Jesper M{\o}ller}

\author{Ege Rubak}

\affil{Department of Mathematical Sciences, Aalborg University,
  Denmark\authorcr jm@math.aau.dk, rubak@math.aau.dk}

\date{}

\maketitle

\begin{abstract}
We study point
  processes on $\mathbb S^d$, the $d$-dimensional unit sphere $\mathbb S^d$,
  considering both the isotropic and the anisotropic case, and focusing mostly on the spherical case $d=2$. The first part studies  
  reduced Palm distributions
  and functional summary statistics, including nearest neighbour
  functions, empty space functions, and Ripley's and inhomogeneous
  $K$-functions. The second part partly
discusses the appealing properties of determinantal point process
  (DPP) models on the sphere and partly considers the application of functional summary statistics to DPPs. In fact DPPs exhibit repulsiveness, but we also use
  them together with certain dependent thinnings when constructing
  point process models on the sphere with aggregation on the large
  scale and regularity on the small scale. We conclude with a discussion on future work on
  statistics for spatial point processes on the sphere.

\end{abstract}

\section{Introduction}

\subsection{Aim and motivation}

How do we construct models and functional summary statistics for spatial point processes on the
$d$-dimensional unit sphere $\mathbb{S}^d\subset\mathbb R^{d+1}$ when their realizations
exhibit aggregation or regularity or perhaps a combination of both? Here $d=1,2$ are
the practically most relevant cases and for specificity we let $d=2$ in this paper, noting that
$\mathbb{S}^2$ apart from
a scaling may be considered as an approximation of planet Earth. 
However, our discussion can easily be extended to point processes on 
$\mathbb{S}^1$ and the general case of $\mathbb{S}^d$ may be covered
as well.

The first part concerns point processes on $\mathbb{S}^2$ (Section~\ref{s:pps}) and in particular Palm distributions and
functional summary statistics (Section~\ref{s:Palmfunc}). In
the isotropic case of a point process on the sphere,
\cite{robeson:li:huang:14} studied an extension of Ripley's $K$-function to the sphere without
providing the mathematical details for the reduced Palm distribution, which they first use in their definition of the $K$-function on the sphere and second relate to the second order intensity (or pair correlation) function without any proof. We
provide this definition without assuming isotropy so that the
inhomogeneous $K$-function, introduced in
\cite{baddeley:moeller:waagepetersen:97} for point processes on
Euclidean spaces, can be defined on the sphere as well. Moreover, in
the isotropic case, we introduce further useful functional summary
functions, namely the nearest neighbour function, the empty space
function, and the related so-called $J$-function. 

\begin{figure}
  \includegraphics[width=\textwidth]{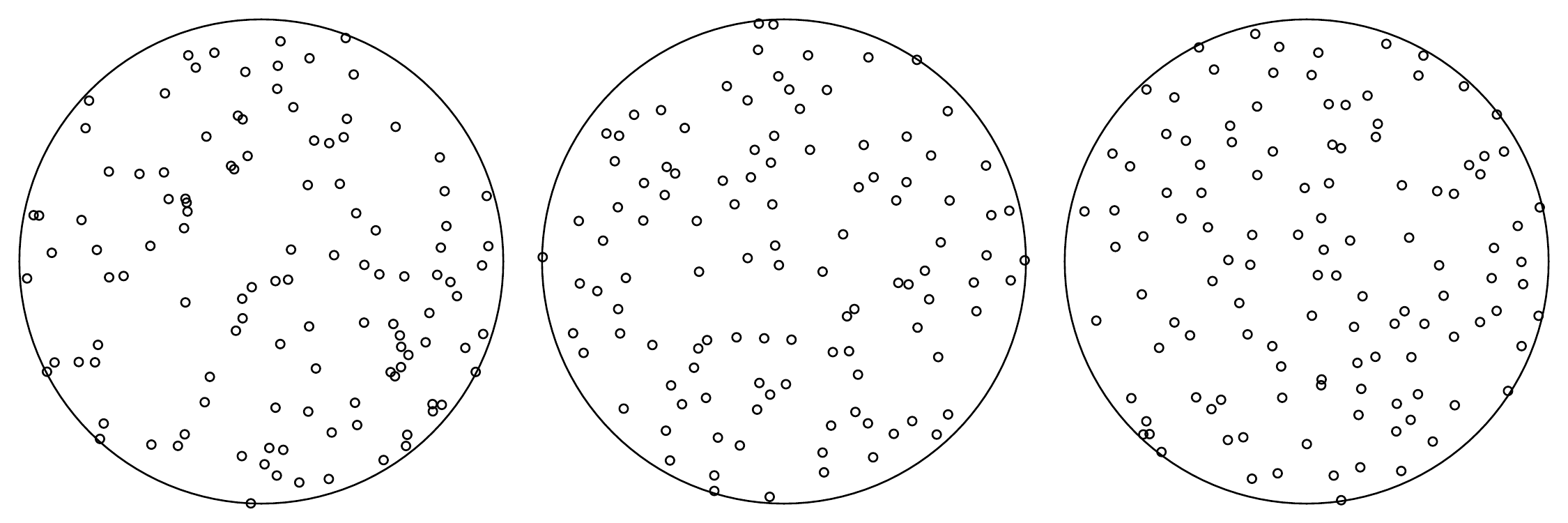}
  \caption{Northern Hemisphere of three spherical point patterns
    projected to the unit disc with an equal-area azimuthal
    projection.  Each pattern is a simulated realization of a
    determinantal point process on the sphere with mean number of
    points 225.  Left: Complete spatial randomness (Poisson process).
    Middle: Multiquadric model with $\tau=10$ and $\delta=\nobreak
    0.68$.  Right: Most repulsive DPP.  }
  \label{fig:realizations}
\end{figure}

The second part partly reviews the attractive properties of determinantal point process
  (DPP) models on the sphere and considers the application of functional summary statistics to DPPs. Briefly, DPPs offer relatively
  flexible models for repulsiveness (although less flexible than Gibbs
  point processes), they can be easily simulated,
  and their moments and the likelihood are tractable, cf.\
  \cite{Hough:etal:09,LMR2,LMR1}. They have mostly been studied and applied for point processes on $\mathbb R^d$, though many results can easily be adapted or may even be simpler for point processes on $\mathbb S^d$, cf.\ \cite{moller:nielsen:porcu:rubak:15}. They are of interest because of
  their applications in mathematical physics, combinatorics,
  random-matrix theory, machine learning, and spatial statistics (see
  \cite{LMR1} and the references therein). Simulated examples of DPPs on the sphere
  with various degrees of
repulsiveness are shown in Figure~\ref{fig:realizations}.

Finally, Section~\ref{s:concluding} contains our concluding remarks, including a discussion
on future work on statistics for spatial point processes on the
sphere.

\subsection{Related work and software}

When we had completed the first version of this paper (see \cite{moeller:rubak:16}) we realized that similar independent research on functional summary statistics by \cite{lawrence:etal:16} was submitted to another journal at the same time as our work, but their paper and our paper supplement each other in different ways:
\cite{lawrence:etal:16} analyze data fitted by a model for clustering (a so-called Thomas process where for the offspring distribution a von Mises-Fisher distribution on the sphere is replacing the bivariate normal distribution used for a planar Thomas process), while we instead consider DPPs which model regularity.
While we provide the details on Palm distributions, they discuss edge correction factors for the functional summary statistics in more detail than we do how.      
Moreover, partly following \cite{lavancier:moeller:16}, we
use DPPs together with certain dependent thinnings when constructing
point process models with aggregation on the large scale and
regularity on the small scale. 
Our paper also provides a survey of and a 
supplement to 
\cite{moller:nielsen:porcu:rubak:15}, which deals with DPPs on $\mathbb S^d$, and where in the present paper we attempt to give a less technical exposition for $d=2$. 

In connection to this paper, the development of software for
simulation of DPPs on the sphere and calculation of non-parametric
estimators for the functional summary statistics constitutes a
substantial amount of work.  The software is written in the \texttt{R}
language \cite{R-core:15} and will be available as an extension of the
\texttt{spatstat} package~\cite{Baddeley:etal:15}.
Section~\ref{s:Palmfunc} shows examples of how to use the software and
the type of plots it can produce.

\section{Point processes on the sphere}\label{s:pps}

Throughout this paper we use a notation and make assumptions as follows. 

Consider a simple finite point process $\bX$ on $\mathbb{S}^2$, i.e.,
we can view $\bX$ as a random finite subset of $\mathbb{S}^2$. Let $N$
be its corresponding counting measure, i.e., $N(A)$ denotes the number
of points in $\bX$ falling in a region $A\subseteq\mathbb{S}^2$. The state space for $X$ is then the set of all finite subsets of $\mathbb{S}^2$ equipped with the $\sigma$-algebra $\mathcal F$ generated by the events $\{N(A)=n\}$ for Borel sets $A\subseteq\mathbb{S}^2$ and integers 
$n=0,1,\ldots$.

For
\begin{equation}\label{e:coordinates}
  \bx=(x_1,x_2,x_3)=(\sin\vartheta\cos\varphi,\sin\vartheta\sin\varphi,\cos\vartheta)\in
  \mathbb{S}^2
\end{equation}
where $\vartheta\in[0,\pi]$ is the polar latitude and
$\varphi\in[0,2\pi)$ is the polar longitude, let
\begin{equation}\label{e:nudef}
  \mathrm d\nu(\bx)=\sin\vartheta\,\mathrm d\varphi\,\mathrm
  d\vartheta
\end{equation}
be the surface measure on $\mathbb{S}^2$. Suppose $\rho:\mathbb{S}^2\mapsto[0,\infty)$ is a Borel function so that $a:=\int_{\mathbb{S}^2}\rho(\bx)\,\mathrm d\nu(\bx)$ is finite. Then $\bX$ is a {\it Poisson process} with intensity function $\rho$ if $N:=N(\mathbb{S}^2)$ is Poisson distributed with parameter $a$, and if conditional on $N=n$, the $n$ points in $\bX$ are independent and each point has density proportional to $\rho$. A Poisson process is the case of no interaction, i.e., it is neither
clustered nor repulsive/inhibitive.

For $n=1,2,\ldots$, we say 
that $\bX$ has {\it $n$th order joint intensity}
$\rho^{(n)}:(\mathbb{S}^2)^n\mapsto[0,\infty)$ with respect to the
$n$-fold product surface measure $\nu^{(n)}=\nu\times\dots\times\nu$ if
for any Borel function $h:(\mathbb{S}^2)^n\mapsto[0,\infty)$,
\begin{equation}\label{e:defrhon}
  \mathrm E\sum_{\bx_1,\ldots,\bx_n\in\bX}^{\not=}h(\bx_1,\ldots,\bx_n)=
  \int h(\bx_1,\ldots,\bx_n)
  \rho^{(n)}(\bx_1,\ldots,\bx_n)\,\mathrm
  d\nu^{(n)}(\bx_1,\ldots,\bx_n)
\end{equation}
and $\int\rho^{(n)}\,\mathrm d\nu^{(n)}<\infty$,  
where the expectation is with respect to $\bX$ and $\not=$ over the
summation sign means that $\bx_1,\ldots,\bx_n$ are pairwise distinct.
Intuitively, if $\bx_1,\ldots,\bx_n$ are pairwise
distinct points on $\mathbb{S}^2$, then
$\rho^{(n)}(\bx_1,\ldots,\bx_n)\,\mathrm
d\nu^{(n)}(\bx_1,\ldots,\bx_n)$ is the probability that $\bX$ has a
point in each of $n$ infinitesimally small regions on $\mathbb{S}^2$
around $\bx_1,\ldots,\bx_n$ and of surface measure $\mathrm
d\nu(\bx_1),\ldots,\mathrm d\nu(\bx_n)$, respectively, cf.\
\eqref{e:defrhon}. 
Note that $\rho^{(n)}$ is uniquely determined except on a
$\nu^{(n)}$-nullset.
In particular, $\rho(\bx)=\rho^{(1)}(\bx)$ is the intensity function
(with respect to surface measure) and we define the {\it pair correlation
  function} by
\begin{equation*} 
  g(\bx,\by):=\frac{\rho^{(2)}(\bx,\by)}{\rho(\bx)\rho(\by)}
  \For{if $\rho(\bx)\rho(\by)>0$}.
\end{equation*} 
The definition of $g(\bx,\by)$ when $\rho(\bx)\rho(\by)=0$ depends on the particular model of $\bX$ and is mainly only of importance when considering plots of $g$: If $\bX$ is a Poisson process with intensity function $\rho$, then $\rho^{(n)}(\bx_1,\ldots,\bx_n)=\rho(\bx_1)\cdots\rho(\bx_n)$ and $g(\bx,\by)=1$ if $\rho(\bx)\rho(\by)>0$, so it is natural to let $g(\bx,\by)=1$ if $\rho(\bx)\rho(\by)=0$. As we shall see later, for DPPs it is natural to make another choice. 

Denote $\mathcal O(3)$ the orthogonal
group, i.e., the set of all $3\times3$ matrices $O$ so that
$OO^\top=O^\top O=I$, where $O^\top$ is the transpose of $O$ and $I$
is the $3\times3$ identity matrix. Set $O\bX=\{O\bx:\bx\in\bX\}$.  
We say that $\bX$ is {\it isotropic/homogeneous} if $O\bX$ is distributed as $\bX$ for all $O\in\mathcal O(3)$.  Then
any point in $\bX$ is uniformly distributed on $\mathbb
S^2$, and the intensity function is equal to a constant called the intensity and which we also denote by $\rho$. 

\section{Palm distributions and functional summary
  statistics}\label{s:Palmfunc}

The most popular functional summary statistic for a stationary point
process on $\mathbb R^2$ with intensity $\rho$ is Ripley's
$K$-function \cite{ripley:76}, where $\rho K(t)$ is interpreted as the
mean number of further points within distance $t$ of a typical point
in the process. The formal definition of $K$ requires Palm measure
theory, and its definition can be extended to the inhomogeneous
$K$-function for so-called second order intensity reweighted
stationary point processes
\cite{baddeley:moeller:waagepetersen:97,coeurjolly:moller:waagepetersen:15}.
The adaption of Ripley's $K$-function to a general isotropic point
process on the sphere is given in \cite{robeson:li:huang:14}, without
explicitly specifying the reduced Palm distribution which is needed in
the precise definition.
In fact $\mathbb S^2$ is a so-called homogeneous space and reduced Palm distributions for homogeneous spaces have been studied in \cite{Rother:Zahle:90,Last:10}, but instead of applying this technical setting we derive the reduced Palm distributions from first principles.

Section~\ref{s:Palm} gives a definition of
the reduced Palm distributions without assuming isotropy and
Section~\ref{s:functional} provides the formal definition of Ripley's $K$-function,
the inhomogeneous $K$-function, and the nearest-neighbour function
on the sphere, together with various useful interpretations and
results for non-parametric estimation. 
Later Section~\ref{s:caseDPP} relates all this to DPPs.

\subsection{Palm distribution for a point process on the
  sphere}\label{s:Palm}

Suppose $\bX$ is a point process on $\mathbb S^2$ with an
integrable intensity function~$\rho$. The so-called Campbell-Mecke
formula gives that for any $\bx\in\mathbb S^2$ there exists a finite
point process $\bX^!_{\bx}$ so that
\begin{equation}\label{e:Palmdef}
  \mathrm E\sum_{\bx\in\bX}h(\bx,\bX\setminus\{\bx\})
  =\int\mathrm E h(\bx,\bX^!_{\bx})\rho(\bx)\,\mathrm d\nu(\bx)
\end{equation}
for any non-negative Borel function $h$. Moreover, the distribution of
$\bX^!_{\bx}$ is unique for $\nu$ almost all $\bx$ with $\rho(\bx)>0$,
and it is called the {\it reduced Palm distribution} at the point
$\bx$. If $\rho(\bx)=0$, this distribution can be chosen to be
arbitrary, since this case will play no role in this paper.
Equation \eqref{e:Palmdef} and the uniqueness result follow from a general result in \cite[Proposition 13.1.IV]{daley:vere-jones:08} (noticing that $\bX^!_{\bx} \cup \{\bx\}$ follows what they call the local Palm distribution).

Intuitively, $\bX^!_{\bx}$ follows the conditional distribution of
$\bX\setminus\{\bx\}$ given that $\bx\in\bX$. This interpretation
follows from \eqref{e:Palmdef} or perhaps more easily by assuming that
$\bX$ has a density $f$ (with respect to the unit rate Poisson process
on $\mathbb S^2$) since, for $\rho(\bx)>0$, $\bX^!_{\bx}$ has density
\begin{equation*}
    f^!_{\bx}(\{\bx_1,\ldots,\bx_n\})=f(\{\bx,\bx_1,\ldots,\bx_n\})/\rho(\bx).
\end{equation*}
For example, for a Poisson point process with intensity function $\rho$,
\begin{equation*}
    f(\{\bx_1, \dots, \bx_n\}) = \exp\left(4\pi - \int_{\mathbb{S}^2}\rho(\bx)\,\mathrm d\nu(\bx)\right) \prod_{i=1}^n \rho(\bx_i)
\end{equation*}
and so it follows that $\bX^!_{\bx}$ is distributed as $\bX$ whenever $\rho(\bx)>0$. Another example is a log Gaussian Cox process (LGCP) $\bX$ on the sphere with underlying Gaussian process $\bY=\{\bY(\bx):\bx\in\mathbb S^2\}$, i.e., when $\bX$ conditional on $\bY$ is a Poisson process with intensity function $\exp(\bY(\bx))$ such that this is almost surely integrable with respect to $\nu$. If $\xi$ and $c$ are the mean and covariance functions of $\bY$, then $\bX_{\bx}^!$ is a LGCP where the underlying Gaussian process has mean function $\xi_{\bx}(\by)=\xi(\by)+c(\bx,\by)$ and the same covariance function $c$. This follows along similar lines as in \cite{coeurjolly:moller:waagepetersen:15b}.

Assuming $\bX$ is isotropic with intensity
$\rho>0$, the situation simplifies as follows.
Denote $\mathcal{SO}(3)$ the 3D
rotation group, i.e., $O\in\mathcal{SO}(3)$ if and only if
$O\in\mathcal O(3)$ and $\det O=1$. \cite{lawrence:etal:16} claim without any proof that for any $\bx\in\mathbb S^2$ and $R\in\mathcal{SO}(3)$, $\bX^!_{R\bx}$ is then distributed as $R \bX^!_{\bx}$. We can easily verify this under a mild condition, namely that the distribution of $\bX$ is given by a density $f$ with respect to the unit rate Poisson process on $\mathbb S^2$ (in other words it is absolutely continuous with respect to this Poisson process).
Then
\begin{align*}
    f^!_{R\bx}(\{\bx_1,\ldots,\bx_n\})
    &= f(\{\bx_1,\ldots,\bx_n,R\bx\})/\rho \\
    &= f(\{R^\top\bx_1,\ldots,R^\top\bx_n,R^\top R\bx\})/\rho \\
    &= f^!_{\bx}(R^\top\{\bx_1,\ldots,\bx_n\})
\end{align*}
where in the second identity we use that $f$ is invariant under rotations since $\bX$ is isotropic, and hence the claim is verified.
So it remains to understand what the distribution of $\bX^!_{\be}$ is for an arbitrary fixed point $\be$ on the sphere; in the sequel we choose this to be the North pole $\be=(0,0,1)$.

For $\bx\in\mathbb
S^2\setminus\{\be\}$, there is a unique
$R_{\bx}\in\mathcal{SO}(3)$ so that $R_{\bx}\be=\bx$ and its
axis of rotation is orthogonal to the great circle on $\mathbb S^2$ which
contains $\bx$ and $\be$.
The axis of rotation is given by $\bu(\bx) = \frac{\be \times \bx}{\norm{\be \times \bx}}$, where $\times$ denotes the cross product in $\R^3$.
According to the right hand rule, let $\theta(\bx)\in(0,2\pi)$ denote the angle for the rotation $R_{\bx}$ around $\bu(\bx)$, i.e., $\theta(\bx)=s(\be,\bx)$ if $0<\theta(\bx)\le\pi$ and $\theta(\bx)=2\pi-s(\be,\bx)$ otherwise,
where $s(\be,\bx)=\arccos(\be\cdot\bx)$ denotes great circle distance from $\be$ to $\bx$ and $\cdot$ is the usual inner product on $\mathbb R^3$.
Then $\bx$ is in one-to-one correspondence to $(\theta(\bx), \bu(\bx))$, and by Rodrigues' rotation formula, for any $\bv\in\R^3$,
\begin{equation*}
    R_{\bx}\bm{v} = \cos\theta(\bx) \bm{v} + \sin\theta(\bx) (\bu(\bx)\times\bm{v}) + (1 - \cos\theta(\bx)) (\bu(\bx) \cdot \bm{v}) \bu(\bx).
\end{equation*}
Furthermore, define $R_{\be}=I$. Now, when $R^\top_{\bx}\bX^!_{\bx}$ and $\bX^!_{\mathbf e}$ are identically distributed it follows from \eqref{e:Palmdef} that 
\begin{equation}\label{e:thirddef}
  \mathrm P\left(\bX^!_{\be}\in F\right)=\frac{1}{\nu(A)\rho}\,\mathrm E\sum_{\bx\in\bX\cap A}1\left[R_{\bx}^\top(\bX\setminus\{\bx\})\in F\right]
\end{equation}
for $F\in\mathcal F$ and an arbitrary Borel set $A\subseteq\mathbb S^2$ with $\nu(A)>0$. The following proposition summarizes our results.

\vspace{0.3cm}

\noindent {\bf Proposition 1.}
Suppose $\bX$ is isotropic with intensity
$\rho>0$ and its distribution is absolutely continuous with respect to the unit rate Poisson process on $\mathbb S^2$. Then for $\nu$ almost all $\bx\in\mathbb S^2$, $\bX^!_{\bx}$ is distributed as $R_{\bx}\bX^!_{\mathbf e}$, where the distribution of $\bX^!_{\mathbf e}$ is given by \eqref{e:thirddef}. So if
$k$ is a non-negative measurable function, then
 for $\nu$ almost all $\bx\in\mathbb S^2$,
\begin{equation}\label{e:result3}
  \mathrm Ek\left(R^\top_{\bx}\bX^!_{\bx}\right)=\mathrm Ek\left(\bX^!_{\mathrm e}\right).
\end{equation}

\vspace{0.3cm}

The absolute continuity condition in Proposition~1 will be satisfied for all point process models we have in mind, but the proposition extends to the general case where absolute continuity is not assumed and $R^\top_{\bx}$ in \eqref{e:thirddef} is replaced by a uniformly distributed rotation given that it maps $\bx$ to $\be$. The proof, which kindly has been provided by Markus Kiderlen, is deferred to Appendix~\ref{s:statPalm}. 

Henceforth, for ease of presentation we ignore that our statements about $\bX^!_{\bx}$ only hold for 
$\nu$ almost all $\bx\in\mathbb S^2$.

\subsection{Functional summary statistics for isotropic and second
  order intensity reweighted isotropic point processes on the
  sphere}\label{s:functional}

\subsubsection{The homogeneous case}

Assume $\bX$ is an isotropic point process on $\mathbb S^2$ with
intensity $\rho>0$ and its
distribution is absolutely continuous with respect to the unit rate Poisson process on $\mathbb S^2$.  Denote geodesic (or orthodromic or great-circle) distance on the sphere by
\[s(\bx,\by)=\arccos(\bx\cdot\by),
  \For{$\bx,\by\in\mathbb{S}^2$}.
\]
Let $s(A,B)=\inf_{\bx\in A,\by\in
  B}s(\bx,\by)$ be the shortest geodesic distance between
$A,B\subset\mathbb S^2$ and define the {\it nearest neighbour
  function} by
\begin{equation*}
  G(t)=\mathrm P\left(s(\bX^!_{\be},\be)\le t\right)=\mathrm P\left(s(\bX^!_{\bx},\bx)\le t\right),
  \For{$0\le t\le\pi,\ \bx\in\mathbb S^2$},
\end{equation*}
where the last identity follows from Proposition~1.
Thus $G$ is the distribution function for the geodesic distance from a typical
point to the nearest other point in $\bX$. Furthermore, for an
arbitrary point $\bx\in\mathbb S^2$, we define the {\it empty space
  function} by
\begin{equation*}
  F(t)=\mathrm P\left(s(\bX,\bx)\le t\right), \For{$0\le t \le\pi$},
\end{equation*}
and following \cite{lieshout:baddeley:96}, we define the $J$-function
by
\begin{equation*}
  J(t)=\frac{1-G(t)}{1-F(t)}\For{for $F(t)<1$}.
\end{equation*}
Since $\bX$ is isotropic, $F$ does not depend on the choice of
$\bx$. If $\bX$ is a homogeneous Poisson process, then
$F_{\mathrm{Pois}}(t)=G_{\mathrm{Pois}}(t)=\exp(-2\pi\rho(1-\cos t))$
and $J_{\mathrm{Pois}}=1$.

Note that for any Borel set $A\subseteq\mathbb S^2$, 
\begin{equation*}
  \rho\nu(A)G(t)=\mathrm E\sum_{\bx\in\bX\cap
    A}1\left[s(\bX\setminus\{\bx\},\bx)\le t\right].
\end{equation*}
Thinking of $A$ as an observation window, this is a useful result when
deriving non-parametric estimates: Let $\mathcal G\subset\mathbb S^2$
be a finite grid of $m>0$ points. If $\bX$ is fully observed on
$\mathbb S^2$, i.e., $A=\mathbb S^2$, then natural estimates are
\begin{equation*}
  \widehat F(t)=\frac{1}{m}\sum_{\bx\in\mathcal G}1\left[s(\bX,\bx)\le
    t\right]
\end{equation*}
and
\begin{equation*}
  \widehat G(t)=\frac{1}{N}
  \sum_{\bx\in\bX}1\left[s(\bX\setminus\{\bx\},\bx)\le t\right]
\end{equation*}
provided $N=N(\mathbb S^2)>0$.  In case the observation window $A$ is
a proper subset of~$\mathbb S^2$, minus sampling may be used: Let
$A_{\ominus t}=\{\bx\in A:s(\mathbb S^2\setminus A,\bx)>t\}$ be the
set of those points in $A$ with geodesic distance at least $t$ to any
point outside $A$. Then minus sampling gives the estimate
\begin{equation*}
  \widehat G(t)=\frac{1}{N(A_{\ominus t})}\sum_{\bx\in\bX\cap
    A_{\ominus t}}1[s(\bX,\bx)\le t]
\end{equation*}
provided $N(A_{\ominus t})>0$.  Moreover, for $\widehat F$ we choose
the grid so that $\mathcal G\subset A_{\ominus t}$.

Now, we define the {\it $K$-function} by
\begin{equation}\label{e:Kdef}
  K(t)=\frac{1}{\rho}\mathrm E\sum_{\bx\in\bX_{\be}^!}1\left[s(\be,\bx)\le t\right]=\frac{1}{\rho}\mathrm E\sum_{\by\in\bX^!_{\bx}}1[s(\bx,\by)\le t],
  \For{$0\le t\le \pi,\ \bx\in\mathbb S^2$},
\end{equation}
where the last equality follows from Proposition~1.
We have that
\begin{equation}\label{e:local}
  \rho K(t)=\mathrm E\sum_{\by\in\bX^!_{\bx}}1[s(\bx,\by)\le t]
\end{equation}
is the mean number of further points within geodesic distance $t$ of a
typical point in the process.  Furthermore, by Proposition~1 and
\eqref{e:Kdef}, we obtain
\begin{equation}\label{e:bbbb}
  \rho^2\nu(A)K(t)=\mathrm E\sum_{\bx\in\bX\cap A}\sum_{\by\in\bX\setminus\{\bx\}}1\left[s(\bx,\by)\le t\right],
\end{equation}
which is another useful formula for deriving non-parametric
estimates. For example, if $\bX$ is fully observed on $\mathbb S^2$,
\begin{equation}\label{e:Kfully}
  \widehat K(t)=\frac{4\pi}{N(N-1)}\sum_{\bx,\by\in\bX}^{\not=}1\left[s(\bx,\by)\le t\right]
\end{equation}
is a natural estimate. If instead the observation window $A$ is a
proper subset of $\mathbb S^2$, minus sampling gives
\begin{equation*}
  \widehat K(t)=\frac{\nu(A)}{N(A)(N(A)-1)}\sum_{\substack{\bx\in\bX\cap
    A \\ \by\in\bX\cap A_{\ominus t}}}^{\not=}1\left[s(\bx,\by)\le
    t\right]
\end{equation*}
provided $N(A)>1$. In the Poisson case $N(A)(N(A)-1)/\nu(A)^2$ is an unbiased estimator for $\rho^2$, but it may be biased for other cases, which may have dramatic consequences for $\widehat K$ as discussed below.

The estimate \eqref{e:Kfully} was also suggested in
\cite{robeson:li:huang:14}, where plots for all values of
$t\in[0,\pi]$ were considered. Apart from the case of Poisson models
we warn against such plots for the following reason. When $\bX$ has
pair correlation function $g_0$, we have
\begin{equation}\label{e:Kg}
  K(t)=2\pi\int^t_0 g_0(s)\sin s\,\mathrm ds,
  \For{$ 0\le t\le\pi$},
\end{equation}
cf.\ \eqref{e:nudef}-\eqref{e:defrhon} 
and \eqref{e:bbbb}.  For an isotropic/homogeneous Poisson process, the
pair correlation function is $g_{\mathrm{Pois}}=1$, so the
$K$-function is $K_{\mathrm{Pois}}(t)=2\pi(1-\cos t)$. Thus using
\eqref{e:Kfully} gives $\widehat K(\pi)=K_{\mathrm{Pois}}(\pi)=4\pi$,
but for non-Poissonian models $\widehat K(\pi)$ may be seriously
biased, since $K$ is an accumulative function of $g_0$, cf.\
\eqref{e:Kg}. For example, if the pair correlation function for $\bX$
is smaller than one (as in the case of a DPP), we may have $\widehat
K(t)\gg K(t)$ for large values of $t$; we illustrate this in
Section~\ref{s:normalization} below. Therefore we recommend only
interpreting plots of $\widehat K(t)$ for smaller values of $t$: If
for smaller or modest values of $t$, $\widehat K(t)$ is below (above)
$K_{\mathrm{Pois}}(t)$, then we interpret this as inhibition or
repulsiveness (aggregation or clustering) between nearby points in
$\bX$. This interpretation is just like in the case of planar point
processes. Incidentally, a second order Taylor approximation around
$t=0$ gives $K_{\mathrm{Pois}}(t)\approx\pi t^2$, where $\pi t^2$ is
the $K$-function for a planar Poisson process. Similar, when
interpreting plots of non-parametric estimates of $F(t),G(t),J(t)$, we
focus on the behavior for small and modest values of $t$.  We refer to
Section~\ref{s:caseDPP} for examples of how to interpret plots of the
functional summary statistics.  In case of the $G$-function as
compared to $G_{\mathrm{Pois}}$, the interpretation is similar to that
of the $K$-function.

\subsubsection{The inhomogeneous case}

Assume $\bX$ is an inhomogeneous point process on $\mathbb S^2$ with
an integrable intensity function $\rho(\bx)$ and an isotropic pair
correlation function, i.e., it is of the form \eqref{e:ggg}. Then, in
accordance with \cite{baddeley:moeller:waagepetersen:97} we say that
$\bX$ is {\it second order intensity reweighted isotropic (or
  pseudo/correlation isotropic)} and define the {\it inhomogeneous
  $K$-function} by \eqref{e:Kg}. So this definition of $K$ is in
accordance with the isotropic case.  In particular, if $\bX$ is a
Poisson process, then it is second order intensity reweighted
isotropic and we still have $K_{\mathrm{Pois}}(t)=2\pi(1-\cos t)$.

In analogy with \eqref{e:bbbb}, we have
\begin{equation}\label{e:111}
  \nu(A)K(t)=\mathrm E\sum_{\bx\in\bX\cap A}\sum_{\by\in\bX\setminus\{\bx\}}
  \frac{1\left[s(\bx,\by)\le t\right]}{\rho(\bx)\rho(\by)}.
\end{equation}
If the intensity function is known or estimated and $\bX$ is fully
observed on $\mathbb S^2$, this suggests the non-parametric estimate
\begin{equation*}
  \widehat K(t)=\sum_{\bx,\by\in\bX}^{\not=}\frac{1\left[s(\bx,\by)\le
      t\right]}{4\pi\rho(\bx)\rho(\by)},
\end{equation*}
cf.\ \cite{baddeley:moeller:waagepetersen:97}.  If instead the
observation window $A$ is a proper subset of $\mathbb S^2$, minus
sampling gives
\begin{equation*}
  \widehat K(t)=\sum_{\bx\in\bX\cap A,\,\by\in\bX\cap A_{\ominus
      t}}^{\not=}\frac{1\left[s(\bx,\by)\le
      t\right]}{\nu(A)\rho(\bx)\rho(\by)}.
\end{equation*}

Finally, by \eqref{e:Palmdef} and \eqref{e:111}, for $\nu$ almost all
$\bx\in\mathbb S^2$ with $\rho(\bx)>0$,
\begin{equation*}
  K(t)=\mathrm E\sum_{\by\in\bX^!_{\bx}}\frac{1[s(\bx,\by)\le
    t]}{\rho(\by)}.
\end{equation*}
Hence, if $\rho(\by)$ is close to $\rho(\bx)$ for $s(\bx,\by)\le t$,
\begin{equation*}
  \rho(\bx)K(t)\approx \mathrm
  E\sum_{\by\in\bX^!_{\bx}}1[s(\bx,\by)\le t],
\end{equation*}
which is a local version of the interpretation of $K(t)$ in the
isotropic case, cf.\ \eqref{e:local}.

\subsubsection{Normalization of $\widehat K$}\label{s:normalization}

The non-parametric estimate given in \eqref{e:Kfully} effectively
corresponds to estimating $\rho^2$ by $N(N-1)/(4\pi)^2$.  As
previously mentioned, this implies $\widehat
K(\pi)=K_{\mathrm{Pois}}(\pi)=4\pi$ making it the natural choice for
Poisson models. For any isotropic point process it follows easily from \eqref{e:bbbb} and the fact that $\mean(N) = 4\pi\rho$ that $K(\pi) = 4\pi + \frac{1}{\rho}\left(\frac{\var(N)}{\mean(N)} - 1\right)$ and therefore $K(\pi) \ge
4\pi-1/\rho$, with equality for models with a fixed number of points
such as a most repulsive DPP (as defined later in Section~\ref{s:charac}).  The lower bound value $K(\pi) = 4\pi -
1/\rho$ would be obtained by the non-parametric estimator if $\rho^2$
was estimated by $(N/(4\pi))^2$ instead such that
\begin{equation}
  \label{e:Kfully2}
  \widehat K(t) = \frac{4\pi}{N^2}\sum_{\bx,\by\in\bX}^{\not=}1\left[s(\bx,\by)\le t\right].
\end{equation}
Thus, in the special case of a model with a non-random number of
points this estimator may be better, but in general as the true model
is unknown we prefer to use \eqref{e:Kfully}.

\begin{figure}
  \includegraphics[width=\textwidth]{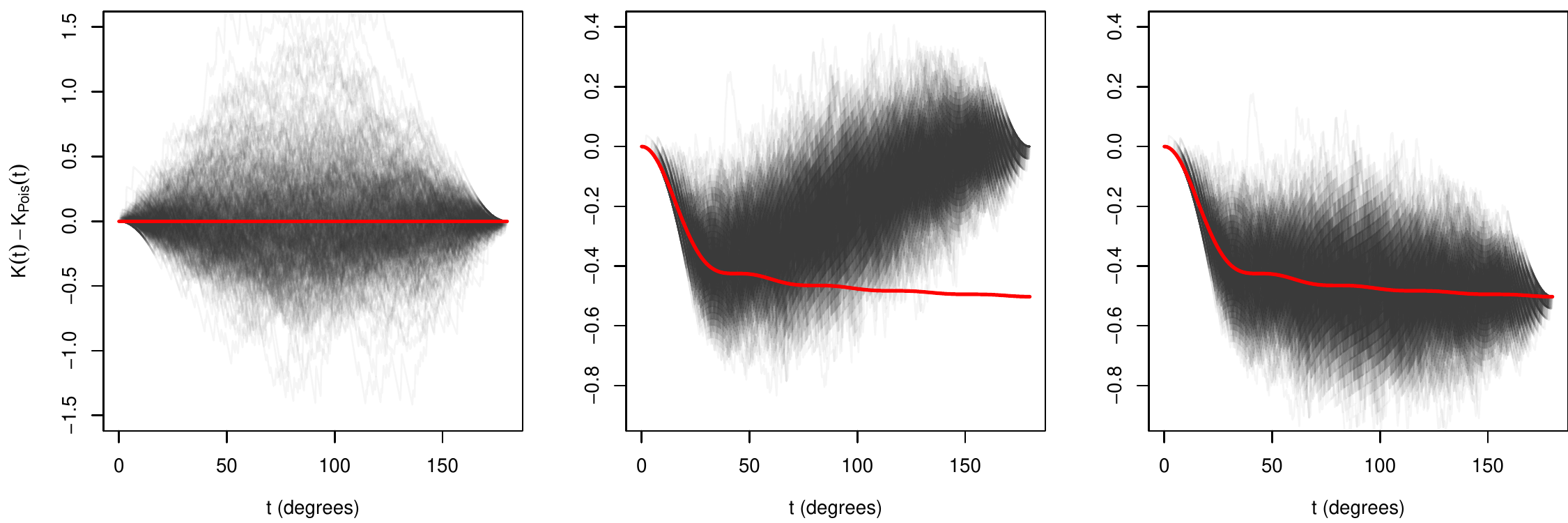}
  \caption{ Each panel shows $\hat K(t) - K_{\textup{Pois}}(t)$ for 500
    simulated point patterns on $S^2$ with 25 points on average
    together with the theoretical value of $K(t) - K_{\textup{Pois}}(t)$
    for the model (red line).  Left: Poisson model and usual
    non-parametric estimator \eqref{e:Kfully}.  Middle: Most repulsive
    DPP and usual non-parametric estimator \eqref{e:Kfully}.  Right:
    Most repulsive DPP and modified non-parametric estimator
    \eqref{e:Kfully2}.  }
  \label{fig:biasKfun}
\end{figure}

Figure~\ref{fig:biasKfun} illustrates the potential bias for a most
repulsive DPP with 25 points (the low number of points helps to
emphasize the bias since the error in this case is $1/\rho =
1/(25/(4\pi)) \approx 0.5$).  Each panel shows the difference between
a non-parametric estimate of $K$ and the theoretical value for a
Poisson process $K_{\textup{Pois}}$ based on 500 simulated point patterns
on $S^2$ under the DPP model.  For reference the left panel shows the perfectly unbiased
result obtained when simulating a Poisson process with 25 points on
average and using \eqref{e:Kfully} to estimate $K$.  The middle panel
shows the bias when \eqref{e:Kfully} is used in the case of a most
repulsive DPP with 25 points.  The right panel shows how the bias is
removed if \eqref{e:Kfully2} is used instead.

In the case of a DPP the bias problem of the non-parametric estimator
for large distances is best illustrated in the somewhat special case
of very low intensity, but we believe the critique and bias problem
remains valid for many other model classes.  In particular we
expect that models for clustering may attain values of $K(\pi)$ much
larger than $4\pi$ and suffer from much larger bias, but it is left as
an open problem to investigate this further.

\section{DPPs on the sphere}\label{s:properties}

We start in Section~\ref{s:def} with the definition of a DPP on
$\mathbb{S}^2$ and  discusses in Section~\ref{s:inhibitive} why a DPP produces
regular point patterns, while the more technical details on existence of a DPP and its 
density function
with respect to a unit rate Poisson process are deferred to 
Appendix~\ref{s:A}. Then in Section~\ref{s:isotropic} we characterize isotropic DPPs and consider parametric models for their kernels. 
Functional summary statistics for such models are discussed in Section~\ref{s:caseDPP}. Finally, in Section~\ref{s:anisotropic} we construct anisotropic DPPs. 

\subsection{Definition and assumptions}
\label{s:def}

Consider again a simple finite point process $\bX$ on $\mathbb{S}^2$. For a
given complex function $C$ defined on the product space
$\mathbb{S}^2\times\mathbb{S}^2$, we say that $\bX$ is a DPP if it has joint intensities 
of any order $n=1,2,\ldots$ which can be expressed in terms of
certain determinants with entries specified by $C$ as detailed
below. An alternative specification in terms of the density for a DPP
is given in Appendix~\ref{s:likelihood}, while Appendix~\ref{s:exists} discusses the technical conditions for the existence of a DPP.  

\vspace{0.3cm}

\noindent {\bf Definition.} $\bX$ is a {\it DPP with kernel $C$} if for all
$n=1,2,\ldots$ and all $\bx_1,\ldots,\bx_n\in\mathbb S^2$,
\begin{equation}\label{e:defdpp}
  \rho^{(n)}(\bx_1,\ldots,\bx_n)=
  \det\left(C(\bx_i,\bx_j)_{i,j=1,\ldots,n}\right)
\end{equation}
where $\det\left(C(\bx_i,\bx_j)_{i,j=1,\ldots,n}\right)$ is the
determinant of the $n\times n$ matrix with $(i,j)$th entry
$C(\bx_i,\bx_j)$. Then we write $\bX\sim\mathrm{DPP}(C)$.

\vspace{0.3cm}

Notice the
following when $\bX\sim\mathrm{DPP}(C)$. 
The intensity function is the diagonal of the kernel:
\begin{equation*}
  \rho(\bx)=C(\bx,\bx), \For{$\bx\in\mathbb S^2$}.
\end{equation*}
The expected number of points is the trace of the kernel:
\begin{equation}\label{e:kappa11}
  \eta:=\mathrm E\left[N\left(\mathbb S^2\right)\right]=\int
  C(\bx,\bx)\,\mathrm d\nu(\bx).
\end{equation}
A Poisson process on $\mathbb S^2$ with intensity function $\rho$ is
the special case of a DPP where $C(\bx,\bx)=\rho(\bx)$ for $\bx \in
\mathbb S^2$, and $C(\bx,\by)=0$ for $\bx\neq\by$. Moreover,
it follows from \eqref{e:defdpp} and since $\rho^{(n)}$ is
non-negative that $C$ has to be positive semi-definite. 

In the remainder of this paper we assume that $\bX\sim\mathrm{DPP}(C)$ where 
 as
in most other works on DPPs, we restrict attention to the case where
the kernel is Hermitian. In other words, $C$ is a complex covariance
function.  We allow the kernel to be complex, since this becomes
convenient when considering simulation of DPPs, cf.\
\cite{moller:nielsen:porcu:rubak:15}. However, isotropy of $C$ implies
that it is real, and all specific models for covariance functions
considered in this paper will be real. Notice that we have already assumed that $C$ is
of finite trace class since it is required that $\eta<\infty$. Finally, we
assume that $C$ is square integrable with respect to $\nu^{(2)}$. 

In summary, we assume that
$C$ is a square integrable complex covariance
function of finite trace class. Then $\mathrm{DPP}(C)$
exists if and only if the spectrum of $C$ is bounded by 0 and 1 (for details, see Appendix~\ref{s:exists}), in which
case $\mathrm{DPP}(C)$ is unique. 

\subsection{Repulsiveness}\label{s:inhibitive}


By~\eqref{e:defdpp} and since $C$ is a covariance function, we have
\begin{equation*}\label{e:compare}
  \rho^{(n)}(\bx_1,\ldots,\bx_n)\le\rho(\bx_1)\cdots\rho(\bx_n)
\end{equation*}
with equality only if $\bX$ is a Poisson process with intensity
function $\rho$. Therefore, a DPP is repulsive unless it is a
Poisson process.
Letting
\begin{equation*}
  R(\bx,\by)=\frac{C(\bx,\by)}{\sqrt{C(\bx,\bx)C(\by,\by)}},
  \For{$\bx,\by\in\mathbb S^2$},
\end{equation*}
be the correlation function corresponding to $C$ when
$\rho(\bx)\rho(\by)>0$, then
\begin{equation*} 
  g(\bx,\by)=1-|R(\bx,\by)|^2
  \For{if $\rho(\bx)\rho(\by)>0$}
\end{equation*} 
and we set $g(\bx,\by)=0$ if $\rho(\bx)\rho(\by)=0$. Thus $g\le1$,
again showing that a DPP is repulsive.

\subsection{Isotropic/homogeneous DPPs}
\label{s:isotropic}

\subsubsection{Characterization of isotropic kernels}\label{s:charac}

Assume that $\bX\sim\mathrm{DPP}(C)$ where the kernel is isotropic,
i.e., it depends only on geodesic distance,
\begin{equation}\label{e:kerneliso}
  C(\bx,\by)=C_0(s),\quad
  s=s(\bx,\by)=\arccos(\bx\cdot\by),
  \For{$\bx,\by\in\mathbb{S}^2$}.
\end{equation}
Further, the
assumption that $C$ is a covariance function implies that $C_0$ is a
real function defined on $[0,\pi]$ such that $C$ is positive
semi-definite.  We follow \cite{DJD_EP13} in calling $C_0$ the {\it
  radial part} of $C$, and we slightly abuse notation and write
$\bX\sim\mathrm{DPP}(C_0)$. 

Clearly, $\bX$ is then an {\it isotropic/homogeneous DPP}.  In
particular any point in $\bX$ is uniformly distributed on $\mathbb
S^2$, the intensity $\rho=C_0(0)$ is constant and equal to the maximal
value of $C_0$, and $\eta=4\pi\rho$ is the expected number of points
in~$\bX$.

In the sequel, assume that $\rho>0$ (otherwise $\bX=\emptyset$). Then
\begin{equation}\label{e:R}
  R_0(s)=C_0(s)/\rho
\end{equation} 
is the radial part of the correlation function $R$ associated
to~$C$. The allowed range of $\rho$ in terms of $R_0$ is the interval
from 0 to
\begin{equation}\label{e:R0}
  \rho_{\max}(R_0)=1/\|R\|
\end{equation}
where $\|R\|<\infty$ denotes the largest eigenvalue of $R$ (see Appendix~\ref{s:exists} and Appendix~\ref{s:merrep}).
Furthermore, the pair correlation function is isotropic and given by
\begin{equation}\label{e:ggg}
  g(\bx,\by)=g_0(s)=1-R_0(s)^2.
\end{equation}
This implies that $g_0(0)=0$ (however, in case of a Poisson process,
it is custom to set $g_0(0)=1$, since $\rho^{(2)}(\bx,\by)=\rho^2$ for
$\nu^{(2)}$ almost all $(\bx,\by)\in\mathbb S^2\times\mathbb S^2$).

Now, assume that $C_0$ is continuous. Then, by a classical result of \cite{MR0005922}, $C_0$ being the radial
part of a continuous isotropic covariance function $C$ is
equivalent to assume that
\begin{equation}\label{e:S}
  C_0(s)=\sum_{\ell=0}^\infty\frac{2\ell+1}{4\pi}\alpha_\ell P_\ell(\cos
  s),\For{$ s\in[0,\pi]$},
\end{equation}
where each $\alpha_\ell\ge0$ is an eigenvalue,
$\sum_{\ell=0}^\infty(2\ell+1)\alpha_\ell<\infty$, and
\begin{equation}\label{s:legendre}
  P_\ell(x)=\frac{1}{2^\ell\ell!}\frac{\mathrm d^\ell}{\mathrm
    dx^\ell}\{(x^2-1)^\ell\},
  \For{$-1<x<1$},
\end{equation}
is the Legendre polynomial of degree $\ell=0,1,\ldots$ (see p.\ 167 in \cite{rainville:60}). 
The eigenvalues $\alpha_\ell$ are also called {\it Mercer coefficients} 
and the collection of Mercer coefficients is
the
  spectrum of the kernel $C$
(see Appendix~\ref{s:exists} and Appendix~\ref{s:merrep}). Note that
  \begin{equation*}
    R_0(s)=\sum_{\ell=0}^\infty\beta_\ell P_\ell(\cos
    s),\For{$ s\in[0,\pi]$},
  \end{equation*}
  where
  \begin{equation*}
    \beta_\ell=(2\ell+1)\alpha_\ell/\eta,
    \For{$\ell=0,1,\ldots,$}
  \end{equation*}
  is a discrete probability distribution, cf.\ \eqref{e:R} and \eqref{e:S}. Conversely, given a
  continuous correlation function $R_0$, i.e., given the sequence
  $\beta_0,\beta_1,\ldots$, \eqref{e:R0} gives the upper bound on the
  expected number of points:
  \begin{equation*}
    \eta_{\max}(R_0)=\inf\{(2\ell+1)/\beta_\ell:\ell=0,1,\ldots\}.
  \end{equation*}
  
In \cite{moller:nielsen:porcu:rubak:15} we quantify both global
  and local repulsiveness in terms of the pair correlation
  function when the intensity is fixed, and we point out that there is
  a {\it trade-off between intensity and the degree of repulsiveness}.
  Loosely speaking the degree of repulsiveness increases as the 
  spectrum of $C$ tends to a step function which for small
  indices $\ell$ is one and for larger indices $\ell$ is zero.
  Therefore, for any integer $m\ge0$, we refer to a DPP with kernel
  \eqref{e:M1} such that $\alpha_\ell=1$ for $\ell\le m$ and
  $\alpha_\ell=0$ for $\ell> m$ as the {\it most repulsive (isotropic)
    DPP} with $\eta=(m+1)^2$ (since in this case
  $\eta=\sum_{\ell=0}^m(2\ell+1)=(m+1)^2$; see
  \cite{moller:nielsen:porcu:rubak:15} for a definition when $\eta$ is
  any positive number).  The Poisson process is another extreme
  obtained when the spectrum tends to zero (but $\rho$ is still
  fixed).  The right panel of Figure~\ref{fig:realizations} shows a
  realization of the most repulsive DPP when $\eta=15^2=225$.

\subsubsection{Examples}\label{s:exsmodels}

Consider the {\it multiquadric family} \cite{Gneiting:2013} given by
\begin{equation*}
  C_0=\rho R_0,\quad
  R_0(s)=\frac{(1-\delta)^{2\tau}}{(1+\delta^2-2\delta\cos s)^\tau},
  \For*{$ 0<\rho\le\rho_{\max}(R_0)$, $\tau>0$, $0<\delta<1$}.
\end{equation*}
As detailed in \cite{moller:nielsen:porcu:rubak:15}, the eigenvalues
$\alpha_\ell$ can easily be calculated numerically, which makes it
possible to simulate realizations from this model.  In
Section~\ref{s:caseDPP} below we furthermore derive a closed form
expression for the $K$-function for the model, which we will use for
statistical tests, and in future work it can also be used for
parameter estimation (as discussed later in
Section~\ref{s:concluding}).  In \cite{moller:nielsen:porcu:rubak:15}
we show that the model is quite flexible and covers the range from no
to intermediate repulsiveness, but in general it does not cover the
most repulsive DPP (only when the expected number of points is very
low).  The middle panel of Figure~\ref{fig:realizations} shows a
realization of a multiquadric model where first we fixed $\eta=225$
and $\tau=10$, and then $\delta=0.68$ was chosen as the smallest value
such that the model is well-defined.  For practical purposes this
corresponds to the most repulsive multiquadric model (the degree of
repulsiveness grows as $\tau$ grows and $\delta$ decreases).  For
comparison a realization of a Poisson process with $\eta=225$ is shown
in the left panel, and it is easy to visually confirm that the
multiquadric model has a higher degree of repulsiveness than the
Poisson process.

In the special case of $\tau=1/2$ we obtain the {\it inverse
  multiquadric family} where $\beta_\ell=\delta^\ell(1-\delta)$
specifies a geometric distribution and
\begin{equation*}
  \alpha_\ell= \eta{\delta^\ell(1-\delta)}/{(2\ell+1)}, \quad
  \eta_{\max}(R_0)={1}/{(1-\delta)}.
\end{equation*}
To notice the trade-off between the intensity and the degree of
repulsiveness, observe that $\eta_{\max}(R_0)$ is a strictly
increasing function of $\delta$ with range $(1,\infty)$, while since
$R_0(s)$ with $s\not=0$ is a strictly decreasing function of $\delta$,
the DPP becomes less repulsive as $\delta$ increases. In the limit as
$\delta\rightarrow1$ we obtain $R_0(s)=0$ corresponding to a Poisson
process; notice that $\alpha_\ell\rightarrow0$ for
$\ell=0,1,\ldots$ On the other hand, as $\delta\rightarrow0$ and if
$\eta=\eta_{\max}(R_0)$ we obtain the most repulsive DPP but
with the mean number of points only equal to 1. As demonstrated in
\cite{moller:nielsen:porcu:rubak:15}, even for $\eta=10$ the DPP is
rather far away from the most repulsive DPP, and for $\eta=100$ it is
rather close to a Poisson process. So the inverse multiquadric family
may be of limited interest except for theoretical considerations.

For the inverse multiquadric model both the kernel and the Mercer
coefficients are expressible on closed form, while in the general
multiquadric model the Mercer coefficients lend themselves to
relatively simple numerical evaluation.  This is a rather unique case,
and in \cite{moller:nielsen:porcu:rubak:15} we consider a number of
other models and conclude that the most useful approach for obtaining
flexible parametric models that cover the full range of possible
repulsiveness for DPPs is a direct modelling of the spectrum. One example of a flexible model is the case
\begin{equation*}\label{e:powerexp}
  \alpha_{\ell}=\frac{1}{1+\beta\exp\left((\ell/\alpha)^\kappa\right)}
  \,,\For{$\ell=0,1,\ldots,$}
\end{equation*}
where $\alpha>0$, $\beta>0$, and $\kappa>0$ are parameters. Since all
$\alpha_{\ell}\in(0,1)$, the DPP is well defined and has a density
specified by \eqref{e:density}, while $\eta$ may be evaluated by
numerical methods. As demonstrated in
\cite{moller:nielsen:porcu:rubak:15}, the model covers a wide range of
repulsive DPPs, including any homogeneous Poisson process and any most
repulsive DPP.

\subsection{Functional summary statistics for isotropic DPPs}\label{s:caseDPP}

Let $\bX$ be an isotropic DPP with an explicit model for the kernel
$C_0(s)=\rho R_0(s)$. Then the pair correlation function is given by
\eqref{e:ggg} and the $K$-function is given by \eqref{e:Kg}.  In
particular for the multiquadric model of Section~\ref{s:exsmodels} we
can easily derive a closed-form expression for the $K$-function using
integration by substitution:
\begin{equation}
  \label{e:Kmultiquadric}
  K_{\textup{mq}}(s) = K_{\textup{Pois}}(s) - 2\pi
  \frac{(1+\delta)(1-\delta)}{2\delta(1-2\tau)}
  \Bigl( \left( \frac{1 + \delta^2 - 2\delta}{1 + \delta^2 - 2\delta
      \cos(s)} \right)^{2\tau-1} - 1 \Bigr), 
  \For*{$\tau\neq1/2$},
\end{equation}
while for the inverse multiquadric family we get
\begin{equation}
  \label{e:Kinversemultiquadric}
  K_{\textup{imq}}(s) = K_{\textup{Pois}}(s) - 2\pi
  \frac{(1+\delta)(1-\delta)}{2\delta} \ln
  \left(\frac{1 + \delta^2 - 2\delta \cos(s)}{1 + \delta^2 - 2\delta}\right),
  \For*{$\tau=1/2$}.
\end{equation}
The formulae \eqref{e:Kmultiquadric}-\eqref{e:Kinversemultiquadric}
also hold for the inhomogeneous $K$-function when $X$ is a correlation
isotropic DPP obtained by independent thinning of an isotropic DPP
with multiquadric kernel.

For DPP models where we have explicit expressions for the Mercer
coefficients but not for the kernel, we can use
\eqref{e:ggg}-\eqref{e:S} to calculate $g_0$ numerically, and then use
numerical integration to calculate the $K$-function, cf.\
\eqref{e:Kg}.  For instance, this approach has to be used both for the
flexible model mentioned in Section~\ref{s:exsmodels} and for the most
repulsive DPP, and it was used to produce the theoretical (red) curve
in the middle and right panels of Figure~\ref{fig:biasKfun}.

The three point patterns in Figure~\ref{fig:realizations} are
realizations of DPPs with different degrees of repulsiveness: From
left to right, there is none (Poisson DPP), intermediate (multiquadric
DPP), and strong (most repulsive DPP) interaction.  In the following
we will discuss the corresponding non-parametric estimates of $K$ and
$G$ and to what extend these can be used to discriminate between the
three cases.

Figure~\ref{fig:KGest} shows the non-parametric estimates of $K$ and
$G$ for all three patterns along with the theoretical curve for a
Poisson process, and as expected the estimates generally have smaller
values for the more repulsive models.  To produce this figure with the
developed software, e.g.\ for the multiquadric DPP, we simulate a
realization of the model and estimate $K$ by using the following
commands:
\begin{verbatim}
  mqmodel <- dppMQ(lambda = 225/(4*pi), delta = 0.68, tau = 10)
  Xmq     <- simulate(mqmodel)
  Kmq     <- Ksphere(Xmq, rmax = 10, angle = TRUE)
  plot(Kmq)
\end{verbatim}

\begin{figure}
  \includegraphics[width=\textwidth]{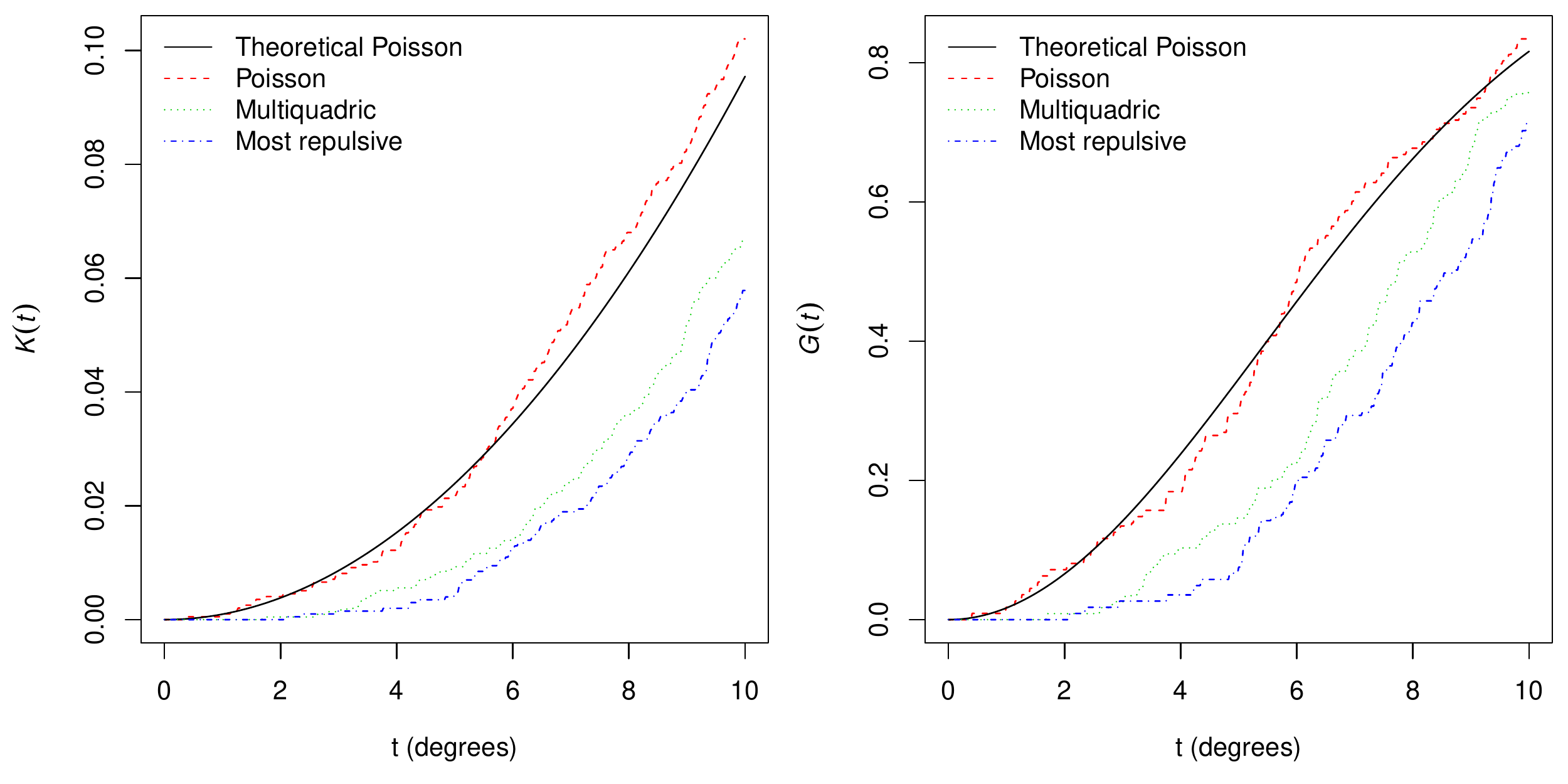}
  \caption{Non-parametric estimates of $K$ (left) and $G$ (right) for
    the three point patterns in Figure~\ref{fig:realizations} and the
    theoretical curves for a Poisson process as reference.  The
    abscissa is the angle in degrees between pairs of points on the
    sphere.  }
  \label{fig:KGest}
\end{figure}

\begin{figure}
  \includegraphics[width=\textwidth]{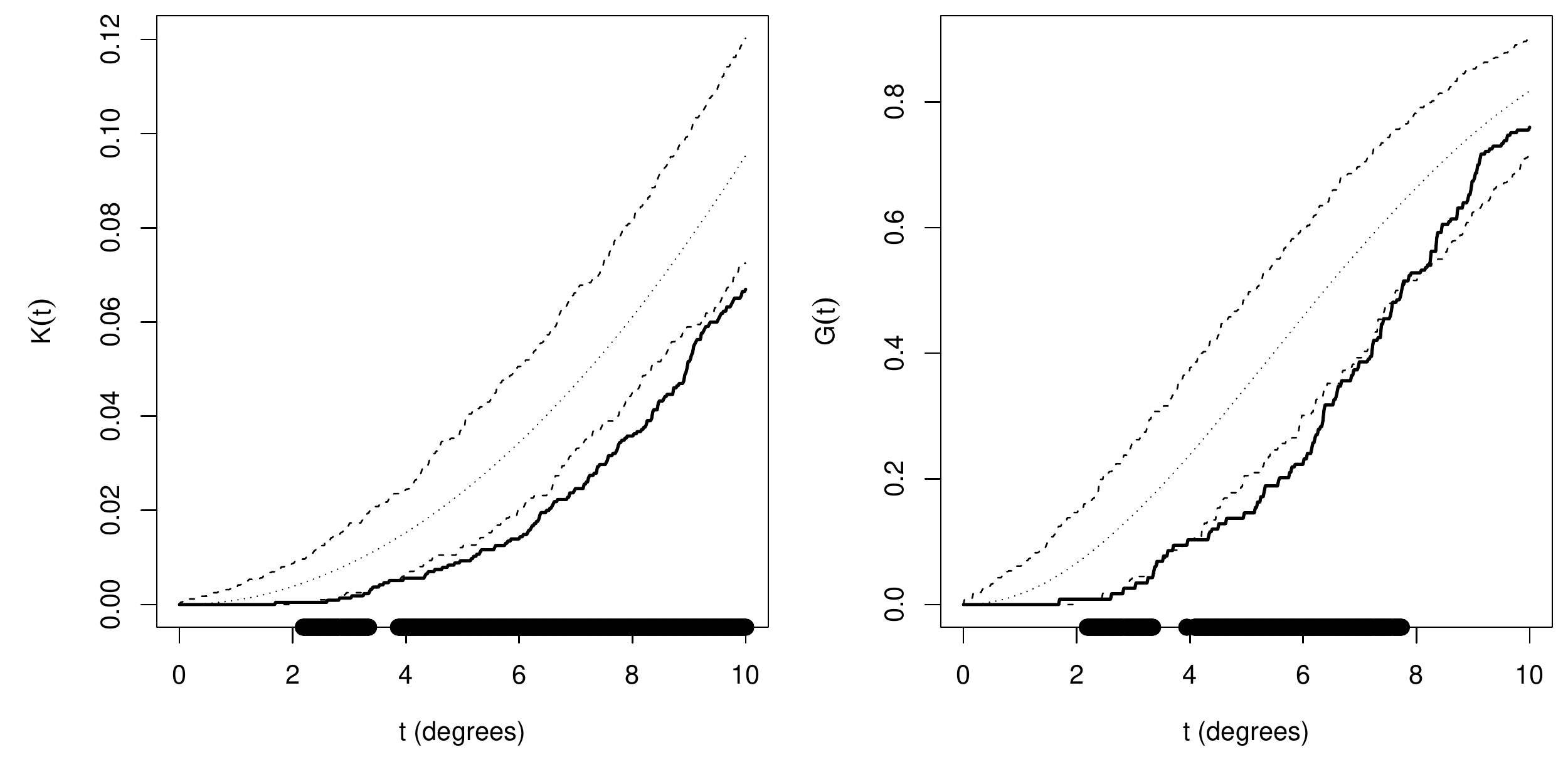}
  \caption{ Simultaneous rank count envelopes for $K$ (left) and $G$
    (right) for the multiquadric point pattern in the middle panel of
    Figure~\ref{fig:realizations} based on 2499 simulations from the
    Poisson null model.  Filled circles on the abscissa correspond to
    distances (in terms of angles in degrees) where the data curve
    exits the envelopes indicating significant departure from the
    Poisson null model.  }
  \label{fig:rank-mqmax2}
\end{figure}

A simple way to assess the difference between the summary statistics
is to use pointwise envelopes simulated under the null model, which is
a technique with a long history for point patterns in Euclidean space
(see e.g.\ \cite{Baddeley:etal:15} for an accessible account).  For
example, in order to generate pointwise envelopes for the
$K$-function, with significance level 1\% for the realization of a
multiquadric DPP generated above under a Poisson null model, and plot
the results (not shown here), we use the commands
\begin{verbatim}
  envmq <- envelope(Xmq, Ksphere, nsim = 199)
  plot(envmq)
\end{verbatim}
This means that if we fix a distance a priori and reject the null
hypothesis if the non-parametric estimate of the summary statistic for
the data is outside the envelopes at this distance, then this is a
test with significance level 1\%.  However, the main drawback is that
in practice it is very hard to only do a pointwise test when the
envelopes show the test results at many scales at once.  This problem
has been well-known for decades and a recent account can be found in
Chapter 10 of \cite{Baddeley:etal:15}.  As an alternative to this
approach so-called rank count envelopes were developed in
\cite{myllymaki:etal:15} which have an interpretation as a global
test, while still providing a graphical output that can be used to
infer the spatial scales where the data significantly deviates from
the null model.  An extra advantage of this approach is that several
functional summary statistics can be combined in one test to give an
overall correct significance level and thereby avoid any multiple
testing problems.  This test was performed on the $K$- and
$G$-function simultaneously for the multiquadric point pattern with a
Poisson null model which based on 2499 simulations yielded a highly
significant $p$-value of 0.0004, which is the lowest possible
$p$-value based on 2500 summary functions (2499 simulated and 1 data).
The corresponding graphical test is shown in
Figure~\ref{fig:rank-mqmax2}, where we have separated the values
related to $K$ and $G$ into separate plots even though the calculation
of the envelopes are based on concatenating the values of $K$ and $G$
into one long vector. Notice that more significant departures from the
null model are detected by the $K$-function which appears to provide a
more powerful test in this case (and in our experience this also
applies to the other examples in this paper).  While it is useful to
know the spatial scales leading to rejection of the null hypothesis,
we should be very cautious when interpreting the $K$-function due to
the cumulative nature of the function, cf.\
Section~\ref{s:functional}.

If we use the same test against the most repulsive DPP as the null
model based on 999 simulations, the $p$-value is 0.001 (which is the
lowest possible value based on 1000 summary functions).  Finally, if
we test the most repulsive pattern (right panel in
Figure~\ref{fig:realizations}) against the multiquadric DPP model with
$\eta=225$, $\delta=0.68$, and $\tau=10$, we get a $p$-value of
$0.008$.  If instead we use $K$ respective $G$ only for the rank count
test, we obtain a $p$-value of $0.005$ respective $0.106$, which again
indicates that $K$ is the more powerful of the two.

\subsection{Models constructed by thinning an isotropic DPP}
\label{s:anisotropic}

This section focuses on {\it anisotropic/inhomogeneous DPP} $\bX$
constructed by independent thinning of an isotropic/homogeneous DPP
$\bY$ on $\mathbb S^2$ with kernel $C_Y$ and $n$th order product
intensity $\rho^{(n)}_Y$. We also follow \cite{lavancier:moeller:16}
in considering a doubly stochastic construction where $\bX$ is
obtained by a dependent thinning of $\bY$. Thereby we can model
regularity on the small scale and clustering on the large scale.

Suppose
\begin{equation*}
  \bX=\{\bx\in\bY: \Pi(\bx)\ge U(\bx)\}
\end{equation*}
where $\bm\Pi=\{\Pi(\bx):\bx\in\mathbb S^2\}$ is a random process
of `selection probabilities' $\Pi(\bx)$, $\bm
U=\{U(\bx):\bx\in\mathbb S^2\}$ is a process of mutually independent
random variables $U(\bx)$ which are uniformly distributed on $[0,1]$,
and $\bY,\bm\Pi,\bm U$ are mutually independent. If
$\bm\Pi$ is deterministic, then $\bX$ is an {\it independent
  thinning} of $\bY$, having $n$th order product intensity
\begin{equation*}
  \rho^{(n)}_X(\bx_1,\ldots,\bx_n)=
  \Pi(\bx_1)\cdots\Pi(\bx_n)\rho^{(n)}_Y(\bx_1,\ldots,\bx_n)
\end{equation*}
and so $\bX$ is seen to be a DPP with kernel
\begin{equation*}
  C_X(\bx,\by)=\Pi(\bx)\Pi(\by)C_Y(\bx,\by).
\end{equation*}
This DPP is anisotropic if $\Pi$ is not constant.
If $\bm\Pi$ is random, then in general $\bX$ is a {\it dependent
  thinning} of $\bY$, with
\begin{equation*}
  \rho^{(n)}_X(\bx_1,\ldots,\bx_n)=
  \mathrm E\left[\Pi(\bx_1)\cdots\Pi(\bx_n)\right]
  \rho^{(n)}_Y(\bx_1,\ldots,\bx_n)
\end{equation*}
and we cannot conclude that $\bX$ is a DPP unless the selection
probabilities are independent.

\begin{figure}
  \includegraphics[width=\textwidth]{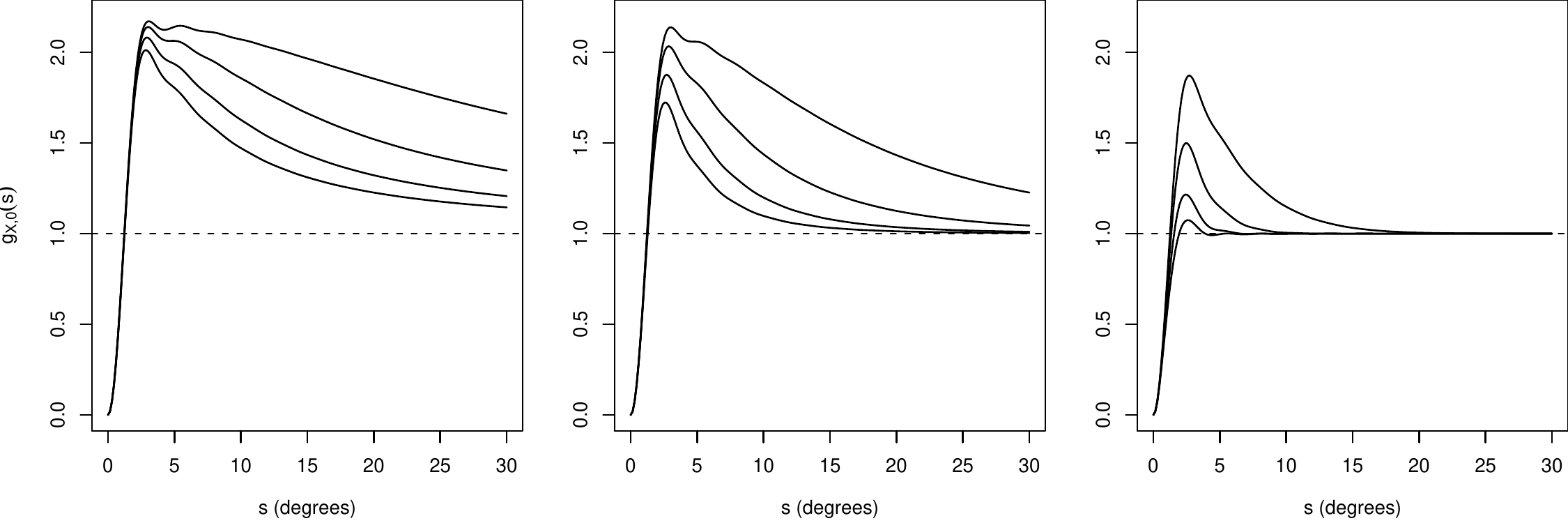}
  \caption{Pair correlation functions for $\Pi$-thinnings of a most
    repulsive DPP with 400 points (see
    Figure~\ref{fig:most-repulsive-400} for a realization).  In all
    panels the covariance function for the underlying Gaussian process
    is multiquadric with variance $\kappa=8$, while $\tau=0.25, 1, 10$
    from left to right panel, and $\delta=0.5, 0.7, 0.8, 0.85$ from
    top to bottom curve within each panel.  }
  \label{fig:pcf-thinned}
\end{figure}

In particular assume that $\bY$ is homogeneous with intensity $\rho_Y$
and pair correlation function $g_{0,Y}(s)$, and the distribution of
$\bm\Pi$ is invariant under the action of $\mathcal O(3)$ on
$\mathbb S^2$. Then $\bX$ is homogeneous, with intensity and pair
correlation function
\begin{equation*}
  \rho_X=q\rho_Y,\qquad g_{X,0}(s)=M_0(s)g_{Y,0}(s),\For{$ s\in[0,\pi]$},
\end{equation*}
where $q=\mathrm E[\Pi(\bx)]$ is the mean selection probability and,
setting $0/0=0$,
\begin{equation*}\label{e:star}
  M_0(s)=M(\bx,\by)=\frac{\mathrm E[\Pi(\bx)\Pi(\by)]}
  {\mathrm E[\Pi(\bx)]\mathrm E[\Pi(\by)]},
  \For{$\bx,\by\in\mathbb S^2$},
\end{equation*} 
depends only on $s=s(\bx,\by)$.  For instance, assume that
$-\log\bm\Pi$ is the $\chi^2$-process given by
\begin{equation}\label{pi-gauss}
  \Pi(\bx)=\exp\left(-\bm Z(\bx)^2/2 \right),
  \For{$\bx\in\mathbb S^2$},
\end{equation} 
where $\bm Z$ is a zero-mean Gaussian process with isotropic
covariance function $K$.  Denoting $K_0$ the radial part of $K$ and assuming the variance $\kappa=K_0(0)$ is positive, we have the same formulas as obtained in \cite{lavancier:moeller:16} but for point processes on $\mathbb R^d$:
\begin{equation*}\label{eq:qkappa}
  q=(1+\kappa)^{-1/2},\quad M_0(s) 
  =\left[1-\frac{R_0(s)^2}{(1+1/\kappa)^2}\right]^{-1/2},
\end{equation*}
where $R_0=K_0/\kappa$ is the (radial part of the) correlation
function of $\bm Z$. Note that $g_{Y,0}\le1$ while $M_0(s)\ge1$ is
typically a decreasing function of $s$. In fact it is possible to
obtain that $g_{X,0}(s)\le1$ for small values of $s$ and
$g_{X,0}(s)\ge1$ for large values of $s$, reflecting regularity on the
small scale and clustering on the large scale.

This is illustrated in Figure~\ref{fig:pcf-thinned}, where the
original process $\bm Y$ is a most repulsive DPP with 400 points
and the underlying Gaussian process $\bm Z$ has a multiquadric
covariance function with variance $\kappa=8$ such that the mean
selection probability is $q=1/3$. Thus the expected number of points
of the thinned process $\bX$ is $400/3$.  As can be seen from
the figure, both $\tau$ and $\delta$ influence the range of the
positive association between points on the longer scale: For both
parameters smaller values yield long range dependence while the
dependence dies out quicker for larger values.  Similar figures (not
shown here) show that changing the original DPP to a multiquadric
model effectively shifts the curves such that the value of $s$ where
$g_{X,0}(s)$ crosses the Poisson reference value 1 shifts to the left,
which is to be expected since the original DPP is less repulsive in
this case.  Note that the geodesic distance in this and subsequent
figures is given in terms of the angle between points on the sphere
measured in degrees from 0 to 180 (as we expect the reader to relate
more easily to these than distances which are effectively in radians).

\begin{figure}
  \includegraphics[width=\textwidth]{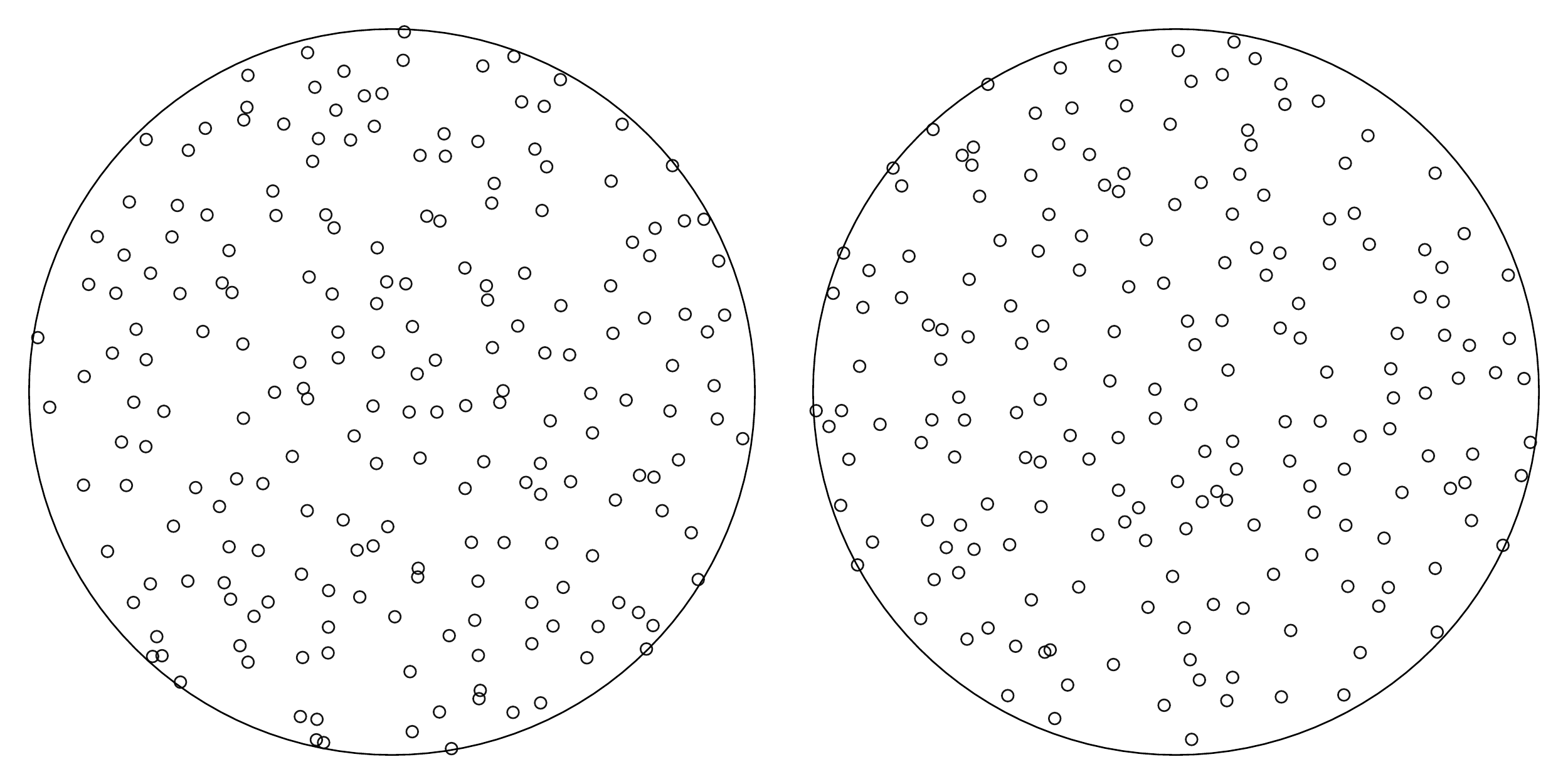}
  \caption{ Northern and Southern Hemispheres of a most repulsive DPP
    with 400 points used as the original process $Y$ before
    $\Pi$-thinning (see text).  The hemispheres are projected to unit
    discs with an equal-area azimuthal projection.  }
  \label{fig:most-repulsive-400}
\end{figure}

\begin{figure}
  \includegraphics[width=\textwidth]{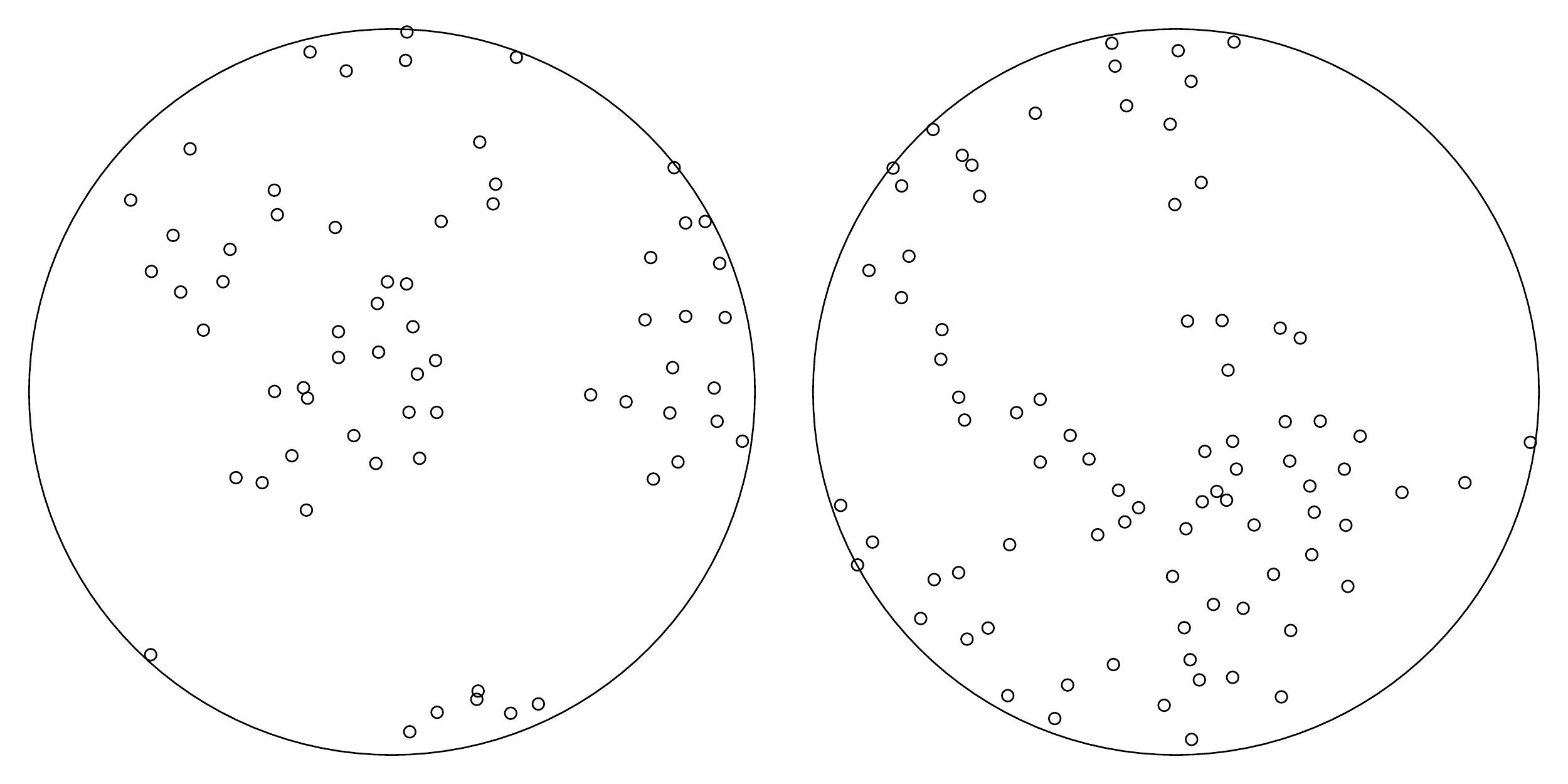}
  \caption{ Northern and Southern Hemispheres of a $\Pi$-thinning of
    the most repulsive DPP with 400 points shown in
    Figure~\ref{fig:most-repulsive-400}.  The Gaussian process
    underlying the $\Pi$-thinning has multiquadric covariance function
    with $\kappa=8$, $\tau=1$, and $\delta=0.5$.  The hemispheres are
    projected to unit discs with an equal-area azimuthal projection
    and the pattern contains 148 points in total.  }
  \label{fig:sim-thinned}
\end{figure}

Note that simulation of $\bZ$ is easy, if we assume that $K$ has
a Mercer representation as in \eqref{e:M1}, with eigenvalues
$\alpha_{\ell}^Z$. Then we generate independent standard normally
distributed random variables $W_{\ell,k}^{(1)}$ and $W_{\ell,k}^{(2)}$
for $\ell=0,1,\ldots$ and $k=-\ell,\ldots,\ell$, and observe that
\begin{equation}\label{e:simZ}
  \sum_{\ell=0}^\infty\sqrt{\alpha_{\ell}^Z}\sum_{k=-\ell}^\ell 
  \left\{W_{\ell,k}^{(1)}\mathrm{Re} \left[Y_{\ell,k}(\bx)\right]
    +W_{\ell,k}^{(2)}\mathrm{Im} \left[Y_{\ell,k}(\bx)\right]\right\},
  \For{$\bx\in\mathbb S^2$},
\end{equation}
is a zero-mean Gaussian process with covariance function $K$ (this
follows from a straightforward calculation, using \eqref{e:M1} and the
fact that $K$ is real).  In practice a truncation of the infinite
series in \eqref{e:simZ} has to be used.  From \eqref{e:S} we have
$\kappa = K_0(0) = \sum_0^\infty \alpha_{\ell}^Z (2\ell+1)/(4\pi)$,
and we choose the truncation such that the truncated series equals
99\% of the given value of $\kappa$.
Figure~\ref{fig:most-repulsive-400} shows a realization of the
original unthinned DPP $\bY$ while Figure~\ref{fig:sim-thinned}
shows the result after $\Pi$-thinning.

Finally, we notice another construction, namely by applying a
one-to-one smooth transformation $\mathbb S^2\mapsto \mathbb S^2$ on
$\bY$ to obtain $\bX$. This results again in that $\bX$ is a DPP with a
kernel that can be specified in terms of $C_Y$ and the derivative of the
transformation. We skip the details here, but see \cite{LMR2,LMR1} for
the result in the case of transformed DPPs on $\mathbb R^d$.

\section{Discussion}\label{s:concluding}

In Section~\ref{s:functional}, for a second order intensity reweighted
isotropic point process on the sphere, we provided non-parametric
estimates of the $F$, $G$ and $K$-functions. In the literature for
point processes defined on Euclidean spaces there is considerable
discussion of edge correction factors, which account for the edge
effects that arise when estimating functional summary statistics near
the boundaries of an observation window.  In
Section~\ref{s:functional}, we exemplified this only in the case when
the process is fully observed or when minus sampling is used, while 
\cite{lawrence:etal:16} provide further edge correction factors.
In the planar case \cite{Baddeley:etal:15} mentions that ``So long as
some kind of edge correction is performed \dots, the particular choice
of edge correction technique is usually not critical.''  We expect the
situation to be similar for point patterns on the sphere.

Our paper started with a brief discussion on how to model aggregation or regularity 
for point processes on the two-dimensional sphere or more generally on $\mathbb S^d$, where two examples are the Thomas process and DPPs. Regarding 
regularity this may
be caused by repulsiveness between the points or by some thinning
mechanism as specified in the following list of models, usually
defined on $\mathbb R^d$ but straightforwardly adapted to
$\mathbb{S}^d$:
\begin{itemize}
\item {\it Mat{\'e}rn hard core processes} of types I-III can be
  simulated by their constructions as dependent thinnings of Poisson
  processes, see
  \cite{huber:wolpert:09,matern:86,moeller:huber:wolpert:10,teichmann:ballani:boogaart:13}.
  However, for the types I-II, the moments of the process will be
  tractable while the likelihood (density) will be intractable; and
  for type III, the opposite is the case.
\item {\it Simple sequential inhibition} and other hard sphere packing
  models (as reviewed in
  \cite{chiu:stoyan:kendall:mecke:13,illian:penttinen:stoyan:stoyan:08})
  can be simulated by their simple constructions of points added one
  by one, but otherwise they are hard to analyse.
\item {\it Gibbs point processes} offer much more flexibility for
  modelling inhibition or repulsiveness, but their simulation may be
  time-consuming and neither the moments nor the likelihood are
  tractable, see
  \cite{moeller:waagepetersen:00,moeller:waagepetersen:06} and the
  references therein.
\end{itemize}
As noticed, DPPs are in many ways more attractive than these models, since DPPs can be easily simulated and their joint intensities and likelihood are tractable. 
In
comparison with DPPs on $\mathbb R^d$, DPPs on $\mathbb{S}^d$ are in many ways
easier to handle, since they are defined on a compact set and we can
more easily deal with the Mercer representation, at least in the isotropic case. 

We have considered examples of simulated point patterns
on the sphere under various DPP models. Indeed it would be interesting
to analyze real point pattern data sets on the sphere using parametric
DPP models. Here we expect that inhomogeneous/anisotropic DPPs will be
of more relevance than homogeneous/isotropic DPPs. As in
\cite{LMR2,LMR1} parameter estimation may be done by either maximum
likelihood or a composite likelihood or minimum contrast method based
on the intensity and pair correlation functions. In \cite{LMR2,LMR1}
we noticed that the latter methods work quite well in comparison with
maximum likelihood.

Other point process models than DPPs on the sphere may of course be of
relevance for applications. For instance, the spectral representation
\eqref{e:spectralrep} allows us to construct and simulate Gaussian
processes, cf.\ \eqref{e:simZ}. Thus we can also deal with log
Gaussian Cox processes (LGCPs) on the sphere, where all the
statistical methodology for LGCPs on Euclidean spaces
\cite{moeller:syversveen:waagepetersen:98,moeller:waagepetersen:00,moeller:waagepetersen:06}
can be easily adapted to the sphere.
 
Finally, we notice that space-time point process models on the sphere,
whether being DPPs or LGCPs or of another type, might be worth
studying, where of course the direction of time should be taken into
consideration.

\subsubsection*{Acknowledgment}
Supported by the Danish Council for Independent Research | Natural
Sciences, grant 12-124675, ``Mathematical and Statistical Analysis of
Spatial Data'', and by the ``Centre for Stochastic Geometry and Advanced
Bioimaging'', funded by grant 8721 from the Villum Foundation. We are grateful to Markus Kiderlen for providing Appendix~\ref{s:statPalm}.

\appendix

\section{Relation between Palm distributions}\label{s:statPalm}

For Borel sets $A\subseteq\mathbb S^2$ and a fixed event $F\in\mathcal F$, denote $\mu_F(A)$ the expected value in the right side of \eqref{e:thirddef}, i.e.,
 \[\mu_F(A)= \mathrm E\sum_{\bx\in\bX\cap A}1\left[R_{\bx}^\top(\bX\setminus\{\bx\})\in F\right].\]
 This is a measure on the Borel sets contained in $\mathbb S^2$, and as verified in Section~\ref{s:Palm} it is rotation invariant if the distribution of $\bX$ is absolutely continuous with respect to the unit rate Poisson process (or any other isotropic Poisson process) on $\mathbb S^2$. However, let $\mathbf e_1,\mathbf e_2,\mathbf e_3$ be the standard basis in $\mathbb R^3$ (so $\mathbf e_3=\mathbf e$) and consider the point process $\bZ=\mathcal O\{\mathbf e_1,\mathbf e_2\}$ where $\mathcal O$ is a uniform rotation (i.e., it follows the normalized Haar measure on $SO(3)$). Obviously, the distribution $\bZ$ is not absolutely continuous with respect to the unit rate Poisson. Further, for $\epsilon>0$ and $\bx\in\mathbb S^2$, denote $A_{\epsilon}(\bx)=\{\by\in\mathbb S^2:s(\bx,\by)<\epsilon\}$ and let $F$ be the set of all finite subsets of $\mathbb S^2$ with at least one point in $A_{2\epsilon}(-\mathbf e_2)$. Then, for $A=A_{\epsilon}(\mathbf e_1)\cup A_{\epsilon}(\mathbf e_2)$,  
 $A'=A_{\epsilon}(\mathbf e_2)\cup A_{\epsilon}(\mathbf e_3)$, and sufficiently small $\epsilon$, it follows straightforwardly that $A'$ is a rotation of $A$, $\mu_F(A)=0$, and $\mu_F(A')>0$. Consequently, $\mu_F$ is not in general rotation invariant, and we need a more complicated construction of the reduced Palm distribution at $\mathbf e$:   
 Let $\theta$ be uniform in $[0,2\pi)$ and denote $\mathcal O_\theta$  the rotation with angle $\theta$ around $\mathbf e$ as the axis of the rotation. Suppose $\theta$ and $\bX$ are independent. It can be shown that for any fixed $F\in\mathcal F$,
\[\chi_F(A)=\mathrm E\sum_{\bx\in\bX\cap A}1[\mathcal O_\theta R_{\bx}^\top (\bX\setminus\{\bx\})\in F]\]
is a rotation invariant measure, and $\mathcal O_\theta R_{\bx}^\top$ follows the distribution of a uniform rotation given that it maps $\bx$ to $\mathbf e$. 
Finally, it follows from a straightforward calculation that $\bX^!_{\bx}$ is distributed as $R_{\bx}\mathcal O_\theta^\top\bX^!_{\mathbf e}$, where 
$\mathrm P\left(\bX^!_{\mathbf e}\in F\right)=\chi_F(A)/[\rho\nu(A)]$ 
 for an arbitrary Borel set $A\subseteq\mathbb S^2$ with $\nu(A)>0$.

\section{Further results for DPPs}\label{s:A}

\subsection{Existence}
\label{s:exists}

Recall that the kernel $C$ is assumed to be a complex covariance function of finite trace class and is square integrable. 
By \cite[Lemma~4.2.6 and Theorem~4.5.5]{Hough:etal:09} $\mathrm{DPP}(C)$
exists if and only if the spectrum of $C$ is bounded by 0 and 1, and
then $\mathrm{DPP}(C)$ is unique. Below we explain what the spectrum is.

Consider a covariance function
$K:\mathbb{S}^2\times\mathbb{S}^2\mapsto\mathbb C$ which is of finite
trace class and is square integrable.  Then, by Mercer's theorem (e.g.\ \cite[Section
98]{riesz:nagy}) and \cite[Lemma~4.2.2]{Hough:etal:09}), ignoring a
$\nu^{(2)}$-nullset, $K$ has a spectral representation
\begin{equation}\label{e:spectralrep}
  K(\bx,\by)=\sum_{n=1}^\infty\alpha_n Y_n(\bx)\overline{Y_n(\by)},
  \For{$\bx,\by\in\mathbb{S}^2$}
\end{equation}
where $Y_1,Y_2,\ldots$ are eigenfunctions which form an orthonormal
basis for the space of square integrable complex functions with
respect to $\nu$.  We call \eqref{e:spectralrep} the {\it Mercer
  representation} of $K$, the eigenvalues $\alpha_i$ for the {\it
  Mercer coefficients}, and
$\operatorname{spec}(K)=\{\alpha_1,\alpha_2,\ldots\}$ the spectrum of $K$.

When $\bX\sim\mathrm{DPP}(C)$, we denote the Mercer coefficients of $C$
by $\lambda_1,\lambda_2,\ldots$, and to ensure existence we require
that $\operatorname{spec}(C)\subset[0,1]$. Then, by \eqref{e:kappa11} and
\eqref{e:spectralrep}, the mean number of points is
$\eta=\sum_{n=1}^\infty\lambda_n$.

\subsection{Likelihood}
\label{s:likelihood}

Suppose $\bX\sim\mathrm{DPP}(C)$ where $\operatorname{spec}(C)\subset[0,1)$. 
Then we can work with the likelihood/density as given below.

Let
$\tilde C: \mathbb{S}^2\times\mathbb{S}^2\mapsto\mathbb C$ be the
complex covariance function with a Mercer representation sharing the
same eigenfunctions as $C$ but with Mercer coefficients
\begin{equation*}\label{e:tildelambda}
  \tilde\lambda_n=\frac{\lambda_n}{1-\lambda_n},
  \For{$ n=1,2,\ldots$}.
\end{equation*} 
Define
\begin{equation*}
  D=\sum_{n=1}^\infty \log\bigl(1+\tilde\lambda_n\bigr).
\end{equation*}
Then, by \cite[Theorem~1.5]{shirai:takahash:03}, $\mathrm{DPP}(C)$ is
absolutely continuous with respect to the unit rate Poisson process
(i.e., the Poisson process on $\mathbb{S}^2$ with intensity measure
$\nu$) and has density
\begin{equation}\label{e:density}
  f(\{\bx_1,\ldots,\bx_n\})=\exp(4\pi-D)
  \det\bigl(\tilde C(\bx_i,\bx_j)_{i,j=1,\ldots,n}\bigr),
\end{equation}
for any point configuration
$\{\bx_1,\ldots,\bx_n\}\subset\mathbb{S}^2$, $n=0,1,\ldots$. 
When
$n=0$ we consider the empty point configuration $\emptyset$, so $\exp(-D)$ is the probability that $\bX=\emptyset$ (since
$\exp(-4\pi)$ is the probability that the unit rate Poisson process is
empty). 

Since
\begin{equation}\label{e:onetoone}
  \lambda_n=\frac{\tilde\lambda_n}{1+\tilde\lambda_n},
  \For{$n=1,2,\ldots$},
\end{equation}
there is a one-to-one correspondence between $C$ and $\tilde C$. Thus,
in order to construct a DPP we can start by specifying any complex
covariance function $\tilde C$ which is of finite trace class and is
square integrable. Its density is then given by \eqref{e:density}, and
$C$ is determined by the Mercer representation of $\tilde C$ and by
\eqref{e:onetoone}. Then $\operatorname{spec}(C)\subset[0,1)$, and so
$\mathrm{DPP}(C)$ is well-defined.

\subsection{Mercer representation for an isotropic kernel}\label{s:merrep}

Consider the isotropic kernel $C$ in \eqref{e:kerneliso}. Below we specify its Mercer representation and discusses when the corresponding DPP exists. 

We need some notation. Recall \eqref{s:legendre} and for
$k=0,\ldots,\ell$ define the associated Legendre functions $P_\ell^{(k)}$
and $P_\ell^{(-k)}$ by (see \cite{dai:xu:13} p.\ 421)
\begin{equation*}
  P_\ell^{(k)}(x) = (-1)^k\left(1-x^2\right)^{k/2}\frac{\mathrm d^k}
  {\mathrm dx^k}P_\ell(x),
  \For{$-1\le x\le 1$},
\end{equation*}
and
\begin{equation*}
  P_\ell ^{(-k)} = (-1)^k \frac{(\ell-k)!}{(\ell+k)!} P_\ell ^{(k)}.
\end{equation*}
Moreover, the surface spherical harmonic functions are given by
\begin{equation*}\label{eq:s_harmonics}
  Y_{\ell,k}(\bx)=Y_{\ell,k}( \vartheta , \varphi ) 
  = \sqrt{\frac{2\ell+1}{4\pi}\,\frac{(\ell-k)!}{(\ell+k)!}} \,
  P_\ell^{(k)} ( \cos{\vartheta} ) \, e^{i k \varphi },
  \For*{$(\vartheta,\varphi)\in[0,\pi]\times[0,2\pi)$},
\end{equation*}
for $k=-\ell,\ldots,\ell$, where $\bx\in\mathbb S^2$ is identified by
its polar latitude and longitude $(\vartheta,\varphi)$, cf.\
\eqref{e:coordinates}, and where $i^2=-1$.  In fact the surface
spherical harmonic functions constitute an orthonormal basis for the
space of square integrable complex functions with respect to $\nu$.

Now, by
the addition formula for spherical harmonics (see
\cite{moller:nielsen:porcu:rubak:15}), \eqref{e:S} is equivalent to the Mercer representation
\begin{equation}\label{e:M1}
  C(\bx_1,\bx_2)=\sum_{\ell=0}^\infty\alpha_\ell\sum_{k=-\ell}^\ell Y_{\ell,k}(\bx_1)
  {\overline{Y_{\ell,k}(\bx_2)}},
  \For{$\bx_1,\bx_2\in\mathbb S^2$}, 
\end{equation}
i.e., the Mercer coefficients are $\lambda_{\ell,k}=\alpha_\ell$, with
$\ell=0,1,\ldots$ and $k=-\ell,\ldots,\ell$. Therefore, to ensure that
$\mathrm{DPP}(C_0)$ is well-defined, we require that the spectrum
$\{\alpha_0,\alpha_1,\ldots\}$ is included in $[0,1]$ and that the sum
\begin{equation*}
  \sum_{\ell=0}^\infty(2\ell+1)\alpha_\ell
\end{equation*}
is finite, and in this case the sum is equal to $\eta$.

 Similarly, if we use the alternative approach of
  Section~\ref{s:likelihood} where we start by specifying $\tilde C$:
  Assuming that $\tilde C$ is a {\it continuous} isotropic covariance
  function is equivalent to that
  \begin{equation}\label{e:M2}
    \tilde C(\bx_1,\bx_2)=\sum_{\ell=0}^\infty\tilde\alpha_\ell Y_{\ell,k}(\bx_1)
    {\overline{Y_{\ell,k}(\bx_2)}},
    \For{$\bx_1,\bx_2\in\mathbb S^2$}, 
  \end{equation}
  where all $\tilde\alpha_\ell$ are non-negative and
  $\sum_{\ell=0}^\infty(2\ell+1)\tilde\alpha_\ell<\infty$.

\bibliographystyle{abbrv}
\bibliography{harmonic,bibliography}

\end{document}